\documentclass[10pt,journal,compsoc]{IEEEtran}

\ifCLASSOPTIONcompsoc
  \usepackage[nocompress]{cite}
\else
  \usepackage{cite}
\fi

\ifCLASSINFOpdf
   \usepackage[pdftex]{graphicx}
\else
\fi

\usepackage{underscore}
\usepackage[document]{ragged2e}
\usepackage[ruled,vlined]{algorithm2e}
\usepackage{dblfloatfix}
\usepackage{caption}
\usepackage{hyperref}
\usepackage[all]{hypcap}
\usepackage[T1]{fontenc}
\DeclareUnicodeCharacter{0301}{\'{e}}

\begin{document}
\title{Decentralizing Supply Chain Anti-Counterfeiting \\ Systems Using Blockchain Technology}
\author{Neo~C.K.~Yiu,~\IEEEmembership{Member,~IEEE\\ Department~of~Computer~Science,~University~of~Oxford\\neo-chungkit.yiu@kellogg.ox.ac.uk}
\IEEEcompsocitemizethanks{\IEEEcompsocthanksitem Neo C.K. Yiu was with Department of Computer Science, University of Oxford, Oxford, United Kingdom}
\thanks{Manuscript first submitted for preprint on Jan 31, 2021.}}

\IEEEtitleabstractindextext{
\justify
\begin{abstract}
An interesting research problem in supply chain industry is evaluating and determining provenance of physical goods – demonstrating authenticity of luxury goods such as bottled wine. Yet, there has been a few innovative software solutions addressing product anti-counterfeiting and record provenance of today's goods that are produced and transported in complex, inter-organizational, and often internationally spanning supply chain networks. However, these supply chain systems and networks have been built and implemented with centralized system architecture, relying on centralized authorities or any form of intermediaries, and leading to issues such as single-point processing, storage and failure, which could be susceptible to malicious modifications of product records or various potential attacks to system components by dishonest participant nodes traversing along the supply chain. Blockchain technology has evolved from being merely a decentralized, distributed and immutable ledger of cryptocurrency transactions to a programmable interactive environment for building decentralized and reliable applications addressing different use cases and existing problems in the world. In this research, the Decentralized NFC-enabled Anti-counterfeiting System (dNAS) is proposed and developed, decentralizing a legacy anti-counterfeiting system of supply chain industry using Blockchain technology, to facilitate trustworthy data provenance retrieval, verification and management, as well as strengthening capability of product anti-counterfeiting in wine industry with capacity to further extend to supply chain industry as a whole. The proposed dNAS utilizes decentralized blockchain network on a consensus protocol compatible with the concept of enterprise consortium, programmable smart contracts and a distributed file storage system to develop a secure and immutable scientific data provenance tracking and management platform on which provenance records, providing compelling properties on data integrity of luxurious goods, are recorded, verified and validated automatically.
\end{abstract}

\begin{IEEEkeywords}
Blockchain, Anti-counterfeiting, Smart contracts, Product Authenticity, End-to-End Traceability, Supply Chain Integrity, Supply Chain Provenance, Decentralization, Enterprise Blockchain, NFC-enabled Anti-counterfeiting System, Near-field Communication, Internet-of-Things, dNAS.
\end{IEEEkeywords}}

\maketitle
\IEEEdisplaynontitleabstractindextext
\justifying
\IEEEpeerreviewmaketitle

\ifCLASSOPTIONcompsoc
\IEEEraisesectionheading{\section{Introduction}\label{sec:introduction}}
\else
\section{Introduction}
\label{sec:introduction}
\fi

\IEEEPARstart{I}{n} this chapter, the motivation of this research and a list of research questions with the research methodology are pinpointed and explained by going through the growing challenges of improving counterfeit product trading. The current solutions which have already been implemented in supply chain industry against the challenges, with the ineffectiveness and inefficiency of current solutions against counterfeit product trading also discussed to better address the need of more innovative and decentralized solutions, such as dNAS, proposed in this research with a set of research objectives declared.

\subsection{The Existing Challenges in Supply Chain Anti-Counterfeiting}
The problem of counterfeit product trading, such as luxurious goods or pharmaceutical products, has been one of the major challenges the supply chain industry has been facing, in an innovation-driven global economy. The situation has exacerbated with an exponential growth of counterfeits and pirated goods worldwide, for which it has also plagued companies with multinational supply chain networks for decades and on.

The analytical study – \cite{1}, published by OECD (Organization for Economic Cooperation and Development) and EUIPO (European Union Intellectual Property Office) in 2016 has further suggested that the volume of counterfeit product trading, which has been alarming, and already amounted to as much as \underline{\emph{\$509 billion}} representing \underline{\emph{3.3\%}} of world trade and 6.8\% of imports from non-EU countries.

In response to the growing concern, innovative, fully-functional, integrable and affordable product anti-counterfeiting solutions, utilizing the cutting-edge technologies, have been widely and urgently demanded so as to ensure provenance and traceability of genuine products throughout supply chain counterfeiting, and suggested solutions should be widely adopted regardless of industries, size of companies and its supply chain systems.

\subsection{Current Alternative Resolutions of Product Anti-Counterfeiting}
Given the growing concern and worsening situation in trading activities of counterfeit products, there have been anti-counterfeiting solutions developed and implemented in the supply chain systems of different industries as listed and explained in \cite{towardblockchain}. Wireless communication technologies, such as \emph{RFID} (Radio-frequency Identification) or \emph{NFC} (Near-field communication which is a subset of RFID), powering the \emph{Internet of Things} (IoT) are mostly those existing anti-counterfeiting and traceability solutions with centralized architectures currently based on, via packaging the tags on packets of goods or products itself for identification and anti-counterfeiting purpose. 

One of the solutions to answer the growing concerns on product counterfeiting along supply chain systems of supply chain industry and specifically for the wine industry, was the \emph{NFC-enabled Anti-Counterfeiting System (NAS)} as depicted in \textit{Appendix~\ref{a1}}. \cite{25} was developed and implemented back in 2014, aiming at providing an innovative and fully functional alternative for supply chain anti-counterfeiting and traceability, based on Near-field communication technology and cloud-based microservice architecture with centralized storage structure, solely hosted by any winemaker, to help improve the worsening situation of product counterfeiting especially for the wine industry.

It has come to a point that even though the implementation of NAS itself is already more secured than most of the typical supply chain systems, with original wine records being validated at any node along the supply chain, centralized architectures of NAS could still pose a risk in data integrity and product authenticity as any node, not only winemakers, along the supply chain have full control of wine records stored in their own database architectures. In case different nodes along the supply chain are untrusting to each other, there could still be possibilities that wine records being duplicated adversely leading to a situation that wine consumers can still purchase bottled wine counterfeits with fabricated wine records retrieved from NAS.

As explained with research findings of security analyses performed in \cite{towardblockchain}, amongst NAS and other existing anti-counterfeiting alternatives with centralized architecture, utilizing wireless tag communication technologies, there are three common counterfeit attacks in the existing centralized anti-counterfeiting and traceability solutions: (1) modification of product records stored in tags, such as fabricating product identifiers or metadata of any wine product, (2) cloning of tag metadata such as those genuine wine records to any counterfeit product tag, and (3) removal of legitimate tags from genuine wine products with reapplications to other counterfeit wine products. 

The typical architecture of centralized supply chains creates several concerns as discussed in \cite{towardblockchain}. First, there is a tremendous processing burden on servers, since significant numbers of products flooded through multiple supply chain nodes. Second, substantial storage is required to store authentication records for every single processed wine product. Third, with existing solutions basing on centralized architecture, traditional supply chains inherently have the problem of single-point failure and so potential service downtime and data loss could be expected. The legacy NAS relies on a centralized authority, such as winemakers, to combat counterfeit products which could in all likelihood result in \emph{single-point processing, storage, and failure}.

\begin{figure}[h]
    \centering
    \captionsetup{justification=centering}
    \includegraphics[width=0.5\textwidth]{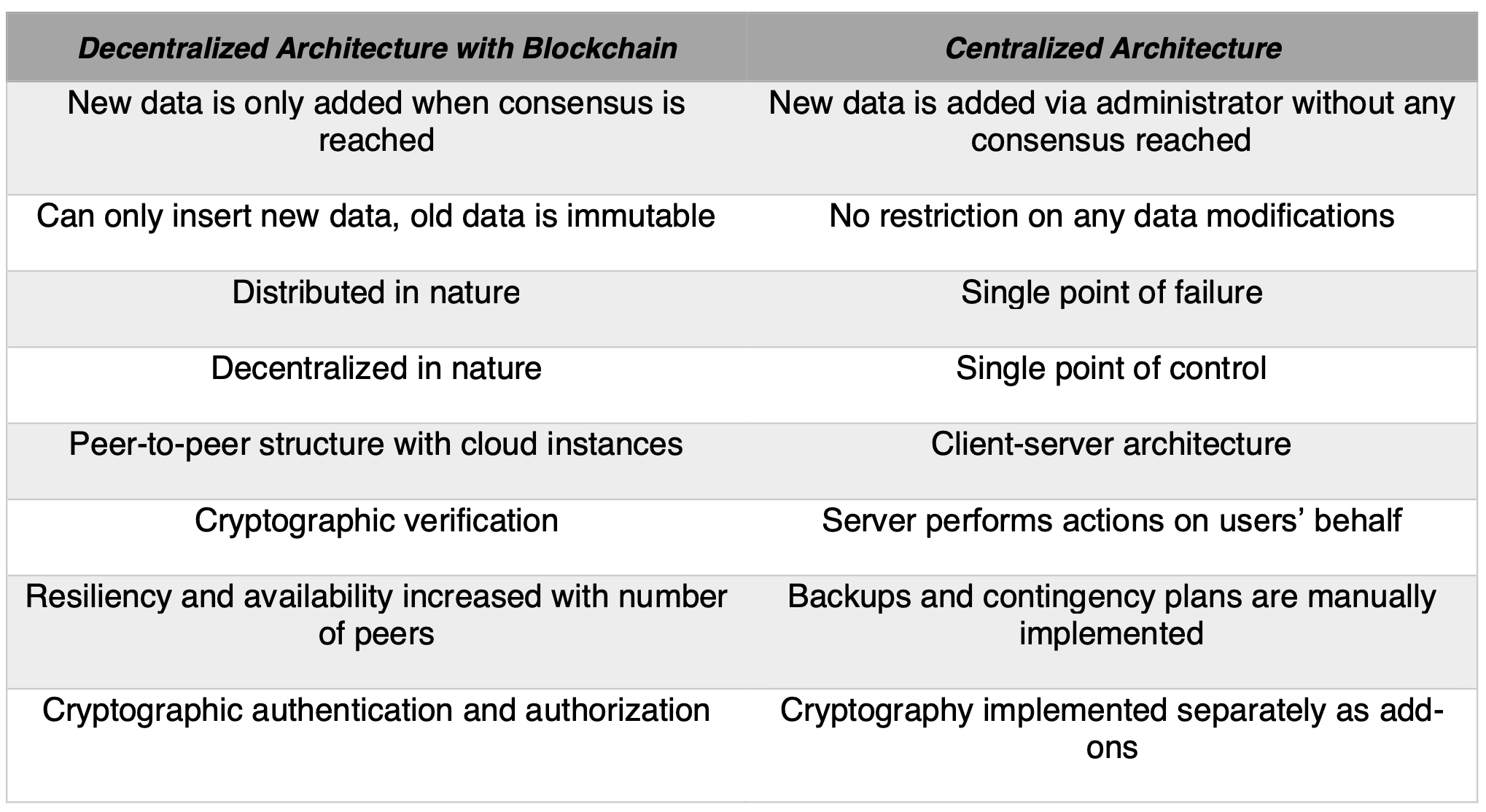}
    \caption{\textit{Comparison of Decentralized and Centralized Architecture}}
    \label{fig:comparisontable}
\end{figure}

To overcome such issues, Blockchain Technology (or other Distributed Ledger Technologies built with decentralized networks) stands out as a potential framework to establish a modernized, decentralized, trustworthy, accountable, transparent, and secured supply chain innovation against counterfeiting attacks, compared with those built with centralized architecture, with comparison between two as detailed in \textit{Fig.~\ref{fig:comparisontable}}. In this research, it is aimed to develop and implement a novel prototype with a decentralized architecture, based on the legacy NAS, to reinforce the innovative idea of product anti-counterfeiting in a decentralized fashion, namely the \emph{Decentralized NFC-based Anti-Counterfeiting System (dNAS)}.

\subsection{Research Objectives}
Open research questions and difficulties inherent in addressing them should first be detailed before actually diving into system design of the proposed dNAS. It will then be followed by outlining a preliminary design of system model and an effective blockchain-based system architecture running on cloud production environment. Given a variety of advantages, such as prevention of single-point failure, better resilience and availability of being applied amongst supply chain participants, brought by the blockchain technology and the concept of decentralized application, to have a more secured and sophisticated supply chain system against counterfeiting attacks, the main question of the research should therefore be:

\emph{"How is it feasible to decentralize an anti-counterfeiting application already developed and implemented in supply chain industry to better combat the rampant counterfeiting attacks?"}

In order to progress the research with a prototype of decentralizing the aforementioned legacy NAS, the main question is then addressed through answering the following sub-questions throughout the research:
\begin{enumerate}
  \item How are the chosen blockchain and its consensus protocol being implemented in the proposed dNAS?
  \item How are the current operations of wine record management and product provenance verification reengineered to be a more decentralized, secured, autonomous and economical alternative?
  \item How is wine record data being stored in the proposed decentralized and distributed storage architecture?
\end{enumerate}

Given the main research question and a set of the derived sub-questions identified, it is common to follow an organised way of exploring them step-wise in this research. The best approach is to follow research techniques applied to the field of computer science namely \emph{proof-by-demonstration} and \emph{implementation-driven} research technique, depicted in \cite{26,27} respectively, under which a working prototype of dNAS will be introduced and validated with the idea of decentralized supply chain solution addressing the main research question and the derived sub-questions. It is important that during development, the next best steps are always taken into account and ultimately offer a better analysis of research questions with detailed documentation of developing such innovative and decentralized solution, using blockchain technology and other useful technical concepts, to improve the exacerbating situation of product counterfeits in supply chain industry.

\section{Background}
In this chapter, following the advantages brought by the blockchain technology, a variety of current blockchain implementations applied to different categories of supply chain industries are also mentioned with reasons on why the proposed dNAS could be a novel and feasible innovation to improve the product counterfeiting, as a primary motive for developing and implementing dNAS. The idea of developing decentralized applications, acting as a contribution of this research for which the development of dNAS is based on, will further be elaborated.

\subsection{Related Work of Blockchain Implementations}
With the advancement of blockchain development in recent years, there have been some existing blockchain innovations developed in different domains and in combination with other emerging technologies for different purposes. A blockchain-based digital certificate system and blockchain-enabled system for personal data protection were proposed in \cite{49,50} respectively. \cite{44,45} have also given an overview of blockchain-based applications developed in different domains where a variety of examples of blockchain systems and use cases are built, could be found in healthcare domain as depicted in \cite{46,47} for decentralized health record management and storage, and energy domain \cite{48} for electric e-mobility, grid management and Peer-to-Peer energy trading.

Blockchain technology has also been utilized to integrate with other technologies, such as \emph{Internet-of-Things} (IoT) and \emph{Artificial Intelligence} (AI), on use cases, such as \cite{51,52} on enabling interoperability of IoT device in a decentralized environment with \emph{Fog computing} come into play as the orchestrating layer, \cite{53} on blockchain-based access control layer to the IoT data storage, and \cite{54} on decentralized AI applications for enhanced data security, improved trust on robotic decisions, decentralized intelligence.

Blockchain innovations have also been implemented across the supply chain industry, and further for the use case of improving product traceability and anti-counterfeiting aspects of the industry. \cite{55} has proposed a concept of blockchain system to enhance transparency, traceability and process integrity of manufacturing supply chains, while an Ethereum-based fully-decentralized traceability system for the Agri-food supply chain management named AgriBlockIoT was developed in \cite{56}. Furthermore, a novel blockchain-based product ownership management system, a blockchain-based anti-counterfeiting system coupled with chemical signatures for additive manufacturing and ontology-driven blockchain design for supply chain provenance, are also detailed in \cite{57,58,59} respectively.

However, the aforementioned blockchain implementations tend to support the findings only with conceptual designs on how blockchain technology could be utilized for achieving better provenance, traceability and process integrity, but few have really been practically implemented with insightful evaluations on the prototypes and strategies regarding the actual implementation and integration in any category of supply chain industry. The issues such as system availability, system integration model, security considerations, data integrity with the actual decentralized data management operations and system functionalities beyond retailer points, are remained to be addressed.

Some implementations are conceptualized and developed based on computation-extensive permission-less blockchain networks and consensus algorithms, aiming at full decentralization over scalability and stability of such decentralized systems developed. Instead of developing blockchain implementations based on conceptual design, decentralizing a legacy anti-counterfeiting system already implemented in the supply chain industry, further with blockchain innovations integrated with, \emph{would be a more pragmatic way to start with} so as to provide timely support to improve the snowballing situation of product counterfeits of supply chain industry, with the novel dNAS developed and implemented.

\subsection{Contribution – Building the Decentralized Application (ÐApp)}
The research is to investigate into the development of decentralised applications (ÐApp) utilizing available frameworks of Enterprise Blockchain, enabling distributed applications to function with unprecedented versatility defined in smart contracts, with a distributed file storage system, such as \emph{InterPlanetary File System} (IPFS) and \emph{Swarm} suggested in \cite{40} and \cite{swarm} respectively, on how it could bring advantages over the legacy solutions in supply chain anti-counterfeiting and traceability. The idea is all about provably honest of which users of the decentralized application should be able to automatically verify the actions requested by the server instances of other nodes along the supply chain, whenever they would like to have the server instances themselves checking with instances of decentralized storage and decentralized network with validation results returned. 

dNAS aims at delivering a more secured and quality approach to verify the authenticity and provenance of luxurious products such as bottled wine, with the use of a peer-to-peer Blockchain Network of Blockchain 2.0 implementations as explained in \cite{towardblockchain}, basing on the concept of enterprise consortium, and a peer-to-peer distributed file storage system, such as IPFS, which has the potential to eliminate absurd amount of costs for on-chain storage and provide enhanced privacy, reliability and quality of service compared with the legacy NAS with centralized architecture. 

Contrary to existing conceptual blockchain implementations applied in different industries, this research is about decentralizing a legacy anti-counterfeiting solution already implemented in supply chain industry, to improve product counterfeits for the industry. \emph{Practicality} in terms of the system implementation, integration model, data integrity, system scalability and stability, is what this research will be after. The proposed prototype depicted in this researach not only stands as a working ÐApps prototype, but also defines a framework and practice for different nodes along the supply chain to integrate the low-cost, real-time and immutable blockchain technology into their daily supply chain workflows. 

For entities to embrace this budding technology with confidence, dNAS needs to be secured, reliable, flexible with integration models available, and its implementation details should be straightforward for the registered supply chain nodes to adopt. Similarly, for consumers interacting with the user interface of dNAS, it needs to be user-friendly and transparent. dNAS itself does offer a foundation to explore strengths and weaknesses of decentralised applications over their centralised counterparts, especially in supply chain industry. The evaluation of the proposed dNAS against the predefined research objectives, and in areas, such as the decentralized wine record management, data and process integrity in dNAS, system security, availability and integration, would also uncover certain limitations and concerns of hosting and deploying ÐApps on Enterprise Blockchain or with other Blockchain technologies and frameworks. The proposed dNAS could further be improved with potential future work inspired from the findings on limitations and concerns raised in the system evaluation process of dNAS.

\section{System Model Definition}
dNAS is a decentralized and integrated system tailor-made for supply chain industry and specifically the wine industry. While capabilities, such as end-to-end traceability, provenance and authenticity were enabled in the legacy NAS, its anti-counterfeiting and traceability model was actually based on winemaker roles with their products moving downstream to their registered supply chain participants and wine consumers until the purchase points. The anti-counterfeiting model of dNAS is actually based on a decentralized \emph{enterprise consortium} consisting of multiple winemakers and supply chain participants, responsible for operating the web application and the NFC-enabled mobile applications of the legacy NAS with a dedicated blockchain network, via an integrated interface, with blockchain nodes assigned to every consortium member. This chapter will go through the use-case analysis, proposed system architecture of dNAS and design decisions made for the development of dNAS.

\subsection{System Requirements of dNAS}
According to the potential opportunities and concerns of developing decentralized solutions for supply chain anti-counterfeiting and traceability, explained in the research discussion result of \cite{towardblockchain}, a fundamental set of system requirements for developing dNAS is therefore defined.

There are in total 19 fundamental system requirements, as listed in \cite{towardblockchain}, to develop dNAS, with rationales ranging from data integrity, scalability, data privacy, improved authentication and authorization, ease of integration, consensus performance to state transparency, summarizing what supply chain industry and even the specific wine industry expected from the innovative dNAS with potential opportunities and concerns of decentralizing supply chain anti-counterfeiting and traceability, also addressed through the development and implementation of dNAS. 

\subsection{System Roles Definition}
Before getting more detailed into the system architecture of dNAS based on the system requirements gathered as discussed, the constituent system roles and their use cases are needed to be clarified. In dNAS, the decentralized architecture and anti-counterfeiting mechanism are building around an enterprise consortium consisted of winemakers and its supply chain participant nodes along the supply chain. Similar to those defined in the legacy NAS, there are mainly FOUR system roles in the proposed system model of dNAS depicted as follows:

\emph{The Winemaker Role}: The winemaker is the manufacturer of wine products. The provided functions of winemakers include creating new wine records with a set of wine pedigree data, supply chain data and transaction data, proposing its supply chain participants to be added to or removed from the consortium as well as its blockchain node to the blockchain network, setting quantity of wine products could be sold, and retrieving information on its supply chain participants so that the latest sales status can be retrieved. The winemaker role is also possible to inquire about wine products that the participants marketed to its wine consumers and verify whether specific wine products have yet been recalled, exchanged and confirmed if the current status of these wine products has yet been verified with public addresses provided by wine consumers. 

\emph{The Supply Chain Participant Role}: The supply chain participants could be with a role of wholesaler, retailer or reseller, and so literally any supply chain node after winemakers and prior to wine consumers. Participants can use functions of dNAS to encrypt wine record components with its private keys, and wine consumers can therefore utilize asymmetric cryptographic concepts applied to system functions in its blockchain service instance hosted by the consortium administrator, with participants' public addresses to verify if the participant is what it claims to be. After each transfer operation, the participant updates the data fields of transaction data and supply chain data, and so the updated content hash is pushed on-chain available for those registered nodes of winemaker role and supply chain participant role to obtain the data required.

\emph{The Wine Consumer Role}: The wine consumers are end purchasers of wine products. The wine consumer can verify whether a supply chain participant has a sales relationship with another winemaker of wine products, and also verify whether the supply chain participant’s inventory has not yet been sold out. Wine consumers can prove that their identities are consistent with their public addresses, and obtain individual transaction records and wine status in wine records via linking the dedicated blockchain nodes to a specific block explorer for reviewing targeted blocks and transactions. As long as a wine product is transferred by a registered node of wine consumer role, the authenticity and end-to-end traceability are still assured and legit even beyond any retailer point post supply chain.

\emph{The Consortium Administrator Role}: A \emph{consortium} is consisted of winemakers and supply chain participants. The consortium administrator is elected through a democratic process involved every consortium member who is assigned with their own blockchain service instance with a designated blockchain node running on the blockchain network managed by the consortium. The proposed dNAS would assume there is only one administrator nodes elected for the \emph{enterprise consortium} to get started in this research for the ease of operations and presentations. The possible approach of having on-chain governance aspects to decentralize the enterprise consortium, or to enhance degree of decentralization on dNAS as a whole, should be covered in future work of implementing dNAS.

\subsection{Use Case Analysis of dNAS}
\begin{figure}[h]
    \centering
    \captionsetup{justification=centering}
    \includegraphics[width=0.4\textwidth]{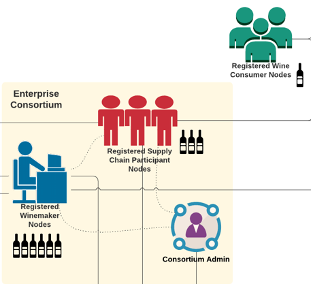}
    \caption{\textit{Enterprise Consortium of dNAS}}
    \label{fig:enterpriseConsortium}
\end{figure}

Given the system roles of dNAS defined, a use case analysis based on these system roles was performed as demonstrated in \textit{Appendix~\ref{b1}}. dNAS is developed based on a concept of \emph{enterprise consortium}, as depicted in \textit{Fig.~\ref{fig:enterpriseConsortium}}, consisted of multiple registered nodes of winemaker role, supply chain participant role and the consortium administrator, introduced to dNAS, unlike the legacy NAS of which the anti-counterfeiting system and mechanism are based \emph{solely on winemakers}.

\begin{figure}[h]
    \centering
    \captionsetup{justification=centering}
    \includegraphics[width=0.48\textwidth]{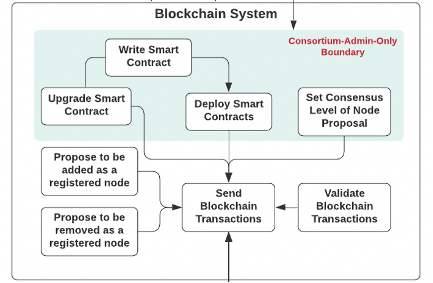}
    \caption{\textit{Use Cases of dNAS Blockchain System}}
    \label{fig:systemusecase}
\end{figure}

Getting started with the blockchain system, as depicted in \textit{Fig.~\ref{fig:systemusecase}}, there are use cases only accessible for the consortium administrator, such as management of smart contracts, which are decentralized software design patterns and methods with access shared across different blockchain nodes running on the network, based on the authorization mechanism defined. The smart contract will need to be developed, deployed or even upgraded to the blockchain network if necessary. The consensus level is indeed the number of votes from consortium members, required for a consensus reached to confirm a registered node to be added to or removed from the consortium. The consensus level can only be set by the consortium administrator if necessary, especially for a situation when the size of the consortium changed drastically affecting degree of decentralization and fraction to reach consensus. For other consortium members, they are able to propose to add a new node to or remove an existing node from the consortium, based on the public address of their designated blockchain node, stored in an on-chain registry. Every consortium member would be assigned with a blockchain node and is therefore eligible to send and validate transactions of the blockchain network to help reach consensus and state broadcasting.

\begin{figure}[h]
    \centering
    \captionsetup{justification=centering}
    \includegraphics[width=0.45\textwidth]{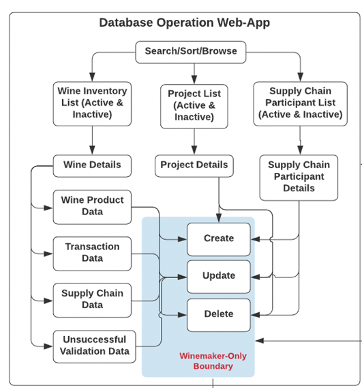}
    \caption{\textit{Use Cases of Database Web Application}}
    \label{fig:databaseusecase}
\end{figure}

Regarding the use cases of the database-operating web application, as listed in \textit{Fig.~\ref{fig:databaseusecase}}, any registered node of winemaker role and supply chain participant role can get access to the functionalities provided by the web application to manage, the inventory of wine products they hold, the list of collaborative supply chain participants, and projects they might work on involving different wine products with different supply chain participants. A wine record, which could be created \emph{only} by registered nodes of winemaker role, is consisted of four constituent data categories such as wine pedigree data, transaction data, supply chain data and unsuccessful validation data. Likewise, only registered nodes of winemaker role could perform operations such as create, update and delete in the web application, while registered supply chain participant nodes and registered wine consumer nodes could only update the respective transaction data and supply chain data via \emph{ScanWINE} mobile application at the point of wine accepting or purchasing following a series of data validation steps, when scanning against the NFC tags on wine products.

\begin{figure}[h]
    \centering
    \captionsetup{justification=centering}
    \includegraphics[width=0.45\textwidth]{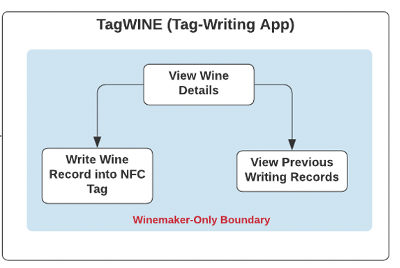}
    \caption{\textit{Use Case of Tag-Writing App - TagWINE}}
    \label{fig:tagwritingusecase}
\end{figure}

Like the legacy NAS, only registered nodes of winemaker role could access to the NFC-enabled tag-writing mobile application - \emph{TagWINE} with use cases as listed in \textit{Fig.~\ref{fig:tagwritingusecase}}, and perform its functionalities such as viewing wine records on the mobile application, writing required data fields of the wine record to NFC tags and checking the history of previous writing activities of those tags on the wine products they produced.

\begin{figure}[h]
    \centering
    \captionsetup{justification=centering}
    \includegraphics[width=0.45\textwidth]{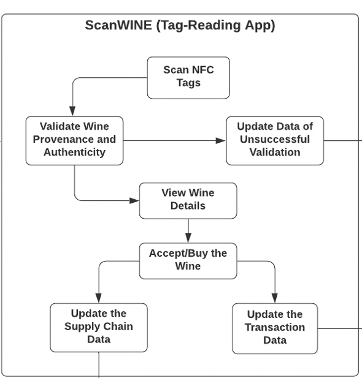}
    \caption{\textit{Use Case of Tag-Reading App - ScanWINE}}
    \label{fig:tagreadingusecase}
\end{figure}

While for the NFC-based tag-reading mobile application - \emph{ScanWINE}, with use cases as listed in \textit{Fig.~\ref{fig:tagreadingusecase}}, any registered nodes, including those of wine consumer role, is accessible to the functionalities, such as scanning NFC tags, which will also trigger the process of validating provenance and authenticity of wine products by analysing the data stored in NFC tags, unique identifier of scanned NFC tags and identity of registered nodes with its registered device scanning the NFC tags. In case such validation process is failed, the unsuccessful validation data of that wine record will be updated and so both related winemaker and supply chain participants would be notified of the new status showed at their database management web application. After going through a successful validation process, the wine record will be shown on the mobile application with further options given, for wine consumers, to buy or, for supply chain participants, to accept the wine product with these physical wine products at their end. The transaction data and the supply chain data are therefore updated following the activities of purchasing or accepting wine products are confirmed.

\begin{figure}[h]
    \centering
    \captionsetup{justification=centering}
    \includegraphics[width=0.45\textwidth]{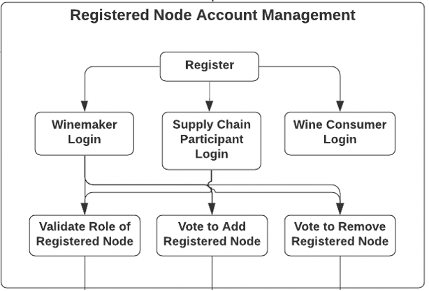}
    \caption{\textit{Use Case of Registered Node Management}}
    \label{fig:nodemanagementusecase}
\end{figure}

Regarding the account management of registered nodes with use cases listed in \textit{Fig.~\ref{fig:nodemanagementusecase}}, it is actually based on the design patterns developed in \emph{AccountController} of the app-backend service in legacy NAS, for which every node could go through registration steps and become registered nodes of either winemaker role, supply chain participant role or wine consumer role. The on-boarding process and registration steps of these aforementioned roles will be different depending on how each type of registered node could get the access to different system components, including the database web application, \emph{TagWINE}, \emph{ScanWINE} and the blockchain network, based on their roles and whether they are members of the \emph{enterprise consortium}. The \emph{AccountController} offers functionalities including validating the role of specific registered nodes, voting to add or remove registered nodes of the \emph{enterprise consortium}.

\section{The Decentralized NFC-Enabled Anti-Counterfeiting System - dNAS}
Following the identified system model definition, a system architecture of dNAS is then designed accordingly, in order to answer the main research question of this research. With the proposed system architecture of dNAS as demonstrated in \textit{Appendix~\ref{b2}}, it explains how exactly the legacy NAS could be re-engineered and running on the chosen blockchain network of enterprise blockchain implementations, such as Ethereum Proof-of-Authority, what exactly are the system components to be decentralized so as to enable core functionalities of the novel dNAS. The proposed system architecture of dNAS can be explained in mainly three parts, namely (1) the decentralized blockchain network, (2) the blockchain interface, and (3) the system components built in the legacy NAS.

\subsection{The Decentralized and Distributed Blockchain Network}
\begin{figure}[h]
    \centering
    \captionsetup{justification=centering}
    \includegraphics[width=0.45\textwidth]{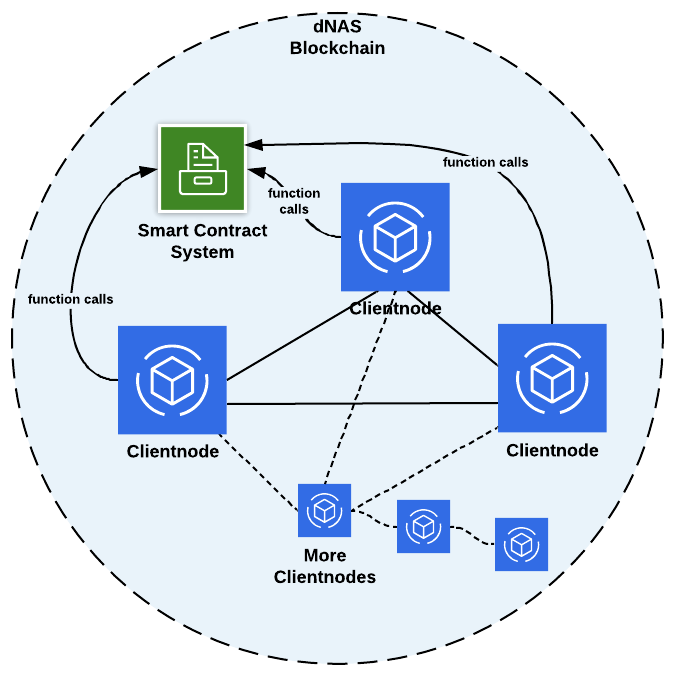}
    \caption{\textit{The dNAS Blockchain Network}}
    \label{fig:dnasblockchain}
\end{figure}

The decentralized blockchain network, as demonstrated in \textit{Fig.~\ref{fig:dnasblockchain}}, is developed based on Enterprise blockchain implmentations such as Quorum blockchain with an consensus algorithm of \emph{Istanbul Byzantine Fault Tolerant} (IBFT) and Ethereum blockchain network with an consensus algorithm of \emph{Proof-of-Authority} (PoA), in line with the concept of enterprise consortium, introduced in the system design and consisted of client nodes which could either be validator nodes (full node) involving in validating transactions and signing new block in the blockchain, or listener node (light node) which could only listen to blockchain states and perform non state-transitioned interactions with smart contracts deployed to the network. Whether a validator node or listener node is assigned will be depending on the role of authorities in the enterprise consortium. For instance, winemaker role would be assigned with a validator node of the blockchain network, while some of the supply chain participant nodes would only be restricted to a listener node depending on its involvement in the enterprise consortium.

The dNAS blockchain network could be built based on enterprise blockchain implementations amongst Quorum blockchain, Ethereum blockchain, Hyperledger Fabric or Tendermint. In case the \emph{Proof-of-Authority} blockchain network of dNAS is developed, it could be built based on different Ethereum client implementations, such as Go-Ethereum, which could specifically termed as Clique Proof-of-Authority network or Aura Proof-of-Authority network. While in case the blockchain network is developed with Quorum blockchain, the \emph{Istanbul Byzantine Fault Tolerant} network could be developed with Quorum client implementations, such as Go-Quorum. No matter which enterprise blockchain implementations the blockchain network of dNAS could develop on, it consists of a set of client nodes managed by different authorities of the enterprise consortium, and thus enabling the peer-permissioning functionalities and so the network is functioned as a permissioned network, unlike public networks such as Ethereum main net which is an open network and not emulated the concept of peer-permissioning, due to the fact that the proposed dNAS is developed for supply chain industry and specifically for the wine industry. When starting client node, individual nodes are connected to each other, via specifying the node identifier of any client nodes already running on the network, so as to form a network.

Smart contracts are developed in Solidity and deployed to the dNAS blockchain network by the consortium administrator who is also hosting validator nodes on the network on behalf of consortium members. Whenever there is a transaction request coming to the blockchain network, the transaction-sending node will invoke methods defined in the smart contract accordingly based on an application binary interface (ABI) specified in data fields of a transaction to perform state-transitioned changes on smart contract storage as well as states of the network with transactions validated and new block mined. The smart contracts can further be upgraded with the implementation of smart contract upgradability pattern. The management of smart contracts with standard processes such as deployment or upgrade is handled by the contract deployment suite for which only blockchain nodes of the consortium administrator will have access to and perform such operations on these smart contracts.

There are also network monitoring components, such as the blockchain explorer and node management suite, to be deployed and connected to blockchain nodes running on the network. The former is a user interface connecting blockchain nodes on the network for authorities to perform querying and viewing on validated transactions and mined blocks including those are at pending stage. The latter is a dashboard listing different performance metrics of every blockchain node running on the network. The data collected from these network monitoring tools could further be augmented and integrated with any open-source data visualization tools, such as Kibana, with enhanced monitoring capability of dNAS as a whole. 

With the advent of the decentralized blockchain network integrated with system components of the legacy NAS via a tailor-made blockchain interface with other tools, the suggested architecture pattern can then provide a high throughput and content-addressed block storage model to the architecture that greatly improves the performance of the proposed dNAS. Similarly, regarding the advantages contributed by adopting Blockchain technology, it has been able to prevent single-point failure, and nodes along the supply chain do not essentially need to trust but collaborate with each other in the enterprise consortium.

\subsection{The Blockchain Interface}
\begin{figure}[h]
    \centering
    \captionsetup{justification=centering}
    \includegraphics[width=0.5\textwidth]{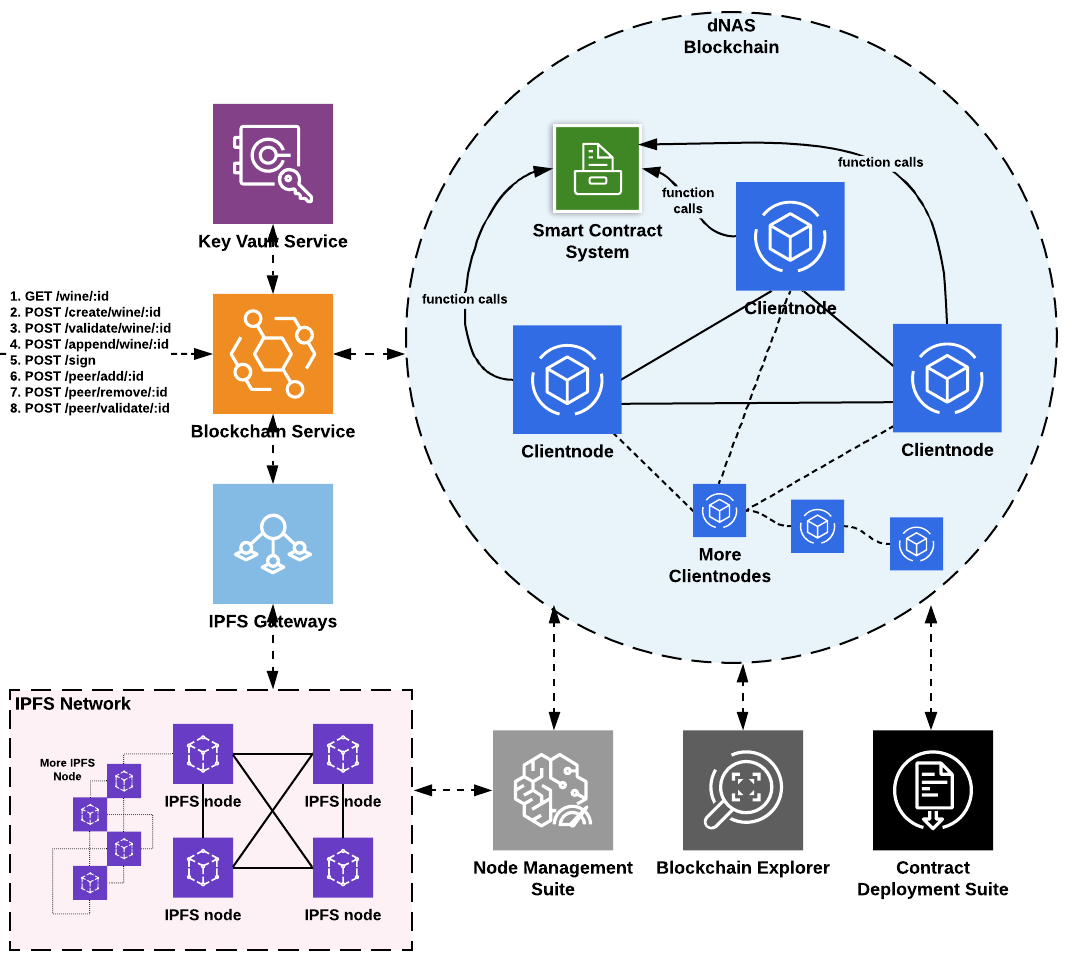}
    \caption{\textit{The Proposed Blockchain Interface of dNAS}}
    \label{fig:dnasblockchaininterface}
\end{figure}

Given the decentralized blockchain network is regarded as a vital part of the proposed dNAS, a dedicated blockchain interface as demonstrated in \textit{Fig.~\ref{fig:dnasblockchaininterface}}, which is essentially another microservice, will need to be designed and developed to help integrate the decentralized bits to system components of the legacy NAS. The blockchain service, acting as an interface between system components of the legacy NAS and the blockchain network, is developed based on functionalities, offered by open-source Ethereum interface libraries such as \emph{Ethers.js} and \emph{web3.js}, aiming to interact with the designated blockchain node and in turn the blockchain network. API endpoints are also developed basing on logic patterns in different controllers of the blockchain service.

The \emph{AppController} of app-backend service in the legacy NAS could invoke endpoints to perform different functionalities, such as pushing data on-chain, getting or validating data from the deployed smart contracts, executing peer-related operations including voting to add or remove participating nodes, and validate roles of registered nodes in the enterprise consortium. The blockchain network is also utilized as a controller of the enterprise consortium managing access control, identities of registered nodes by implementing a concept of on-chain peer registry and serving as a tamper-proof log of transaction events emitted back to the blockchain service. 

With different helper functions developed in the blockchain service and functionalities enabled with open-source Ethereum interface libraries, the blockchain service instance can therefore direct its designated blockchain node, to send transactions so as to perform different operations related to wine records and the peer registry. Sending transactions will require a signature produced with a private key of the node account. Therefore, a dedicated key management module such as the key vault service, will need to be in place to store key pairs related in any signing activities involved in any transaction processed by a specific node account, which is also linked with the blockchain service instance, and to provide a secured mechanism for any key pair to be retrieved whenever there is a transaction or a signing process to be handled.

Instead of storing the complete version of wine record with all of its constituent metadata such as transaction data, wine pedigree data and supply chain data, only the content hash of the wine record along with other payloads for data security and privacy should be stored in the data structure defined in the deployed smart contracts. Distributed data storage system, such as IPFS, is therefore introduced and migrated to the proposed dNAS, so that the dedicated and distributed IPFS node could be connected to an instance of the blockchain service whenever there is a request sending to the blockchain service to process. Storing wine records, in such a distributed storage system, provides a unique content hash which would then return to the blockchain service for further operations before being included in any transaction which would then be signed by dedicated node accounts and sent to the transaction pool of the blockchain network queuing to be processed for validation. 

The content hash of specific wine record is then retrieved from the smart contract whenever there is a validation process to be executed and be further used to retrieve wine record subsets from the distributed storage system off-chain, via the designated IPFS node. IPFS speaks about an idea of a permanent web, implying any data published to the network should be available forever. The node management suite is also connected to IPFS nodes so as to have the same monitoring feature applied to every distributed IPFS node assigned to registered nodes of the enterprise consortium.

\subsection{The Legacy NAS Components}
\begin{figure}[h]
    \centering
    \captionsetup{justification=centering}
    \includegraphics[width=0.48\textwidth]{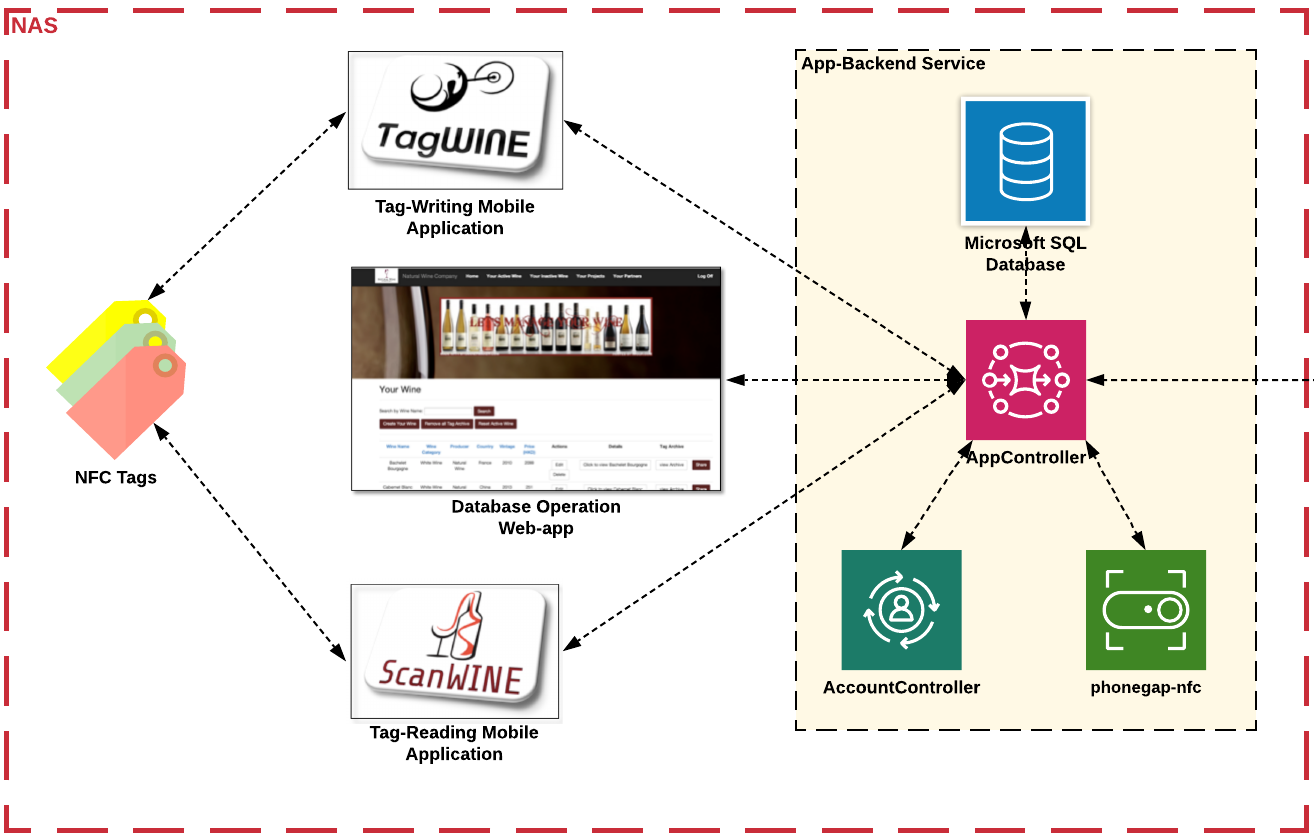}
    \caption{\textit{The System Components of Legacy NAS}}
    \label{fig:legacynas}
\end{figure}

According to system components of the legacy NAS as demonstrated in \textit{Fig.~\ref{fig:legacynas}}, the applications provided to different registered nodes along the supply chain are namely the database-operating web application, \emph{TagWINE} – the NFC-enable tag-writing mobile application, and \emph{ScanWINE} – the NFC-enabled tag-reading mobile application. The mobile applications are designed to interact with the \emph{NTAG 203} of NFC tag model.

The functionalities, enabled by those system components of the legacy NAS, substantiating the use cases described in the use case analysis, are actually based on logic patterns in \emph{AppController} of the app-backend service. For instance, \emph{CRUD} (Create, Read, Update and Delete) operations on wine records, performed on the database operation web application, actually send request to \emph{AppController} invoking respective methods related to such operations, and collaborate with different system components, such as the database system developed in Microsoft SQL, to reflect states transitioned back to the web application. Another example could be using \emph{ScanWINE} to scan a NFC tags, and validate provenance and authenticity of wine products, by validating and comparing respective states of specific wine records stored in the NFC tags, the off-chain database system connected with \emph{AppController}, distributed IPFS storage, and the on-chain data storage of smart contract deployed to dNAS blockchain network via the integration between the app-backend service and the blockchain service.

Regarding components of the app-backend service, a database system is integrated with \emph{AppController} to store states and versions of wine records, and \emph{AccountController} is to process account management of registered nodes with states and user sessions also stored in the database. The \emph{phonegap-nfc} module is integrated with \emph{AppController} to enable \emph{ScanWINE} and \emph{TagWINE} mobile applications with functionalities of any NFC-related interaction, with the specific NFC tags packaged on wine products. Further logic patterns are developed and included in \emph{AppController} to integrate with the decentralized bits of the dNAS via the blockchain interface by invoking related API endpoints of the proposed blockchain service to perform functionalities, such as pushing data to the blockchain and validating on-chain data of specific wine records with transaction events returned to the app-backend service.

\emph{NTAG 203 Ferrite} was the selected type of NFC tag when developing the legacy NAS, given its universal compatibility to most of the NFC-enabled devices and its resistance against metallic interference for which the \emph{NTAG 203 Ferrite Tag} can change the magnetic flux path to avoid interference to the NFC tag, with high surface resistance and effectiveness in preventing resonance and suppressing coupling, from the metallic interference as NFC tags are likely to be attached to foil packaging materials surrounding wine bottlenecks. However, \emph{NTAG 203} chip has already been out of production and replaced by those of \emph{NTAG21x} series, such as \emph{NTAG 213} or \emph{NTAG 216}, simply given the increased memory storage and security functions including password lock and scan counter. \emph{NTAG 216 Ferrite} is therefore the selected type of NFC tag for the development of dNAS owing to its extensive configuration with \emph{7-byte UIDs} and \emph{888-byte memory capacity} for more flexibility in terms of data stored in each NFC tag representing for every unique wine product.

\section{System Implementation}
Given the system model design and system architecture explained, the actual system implementation steps of dNAS are expected. This chapter includes the proposed system implementation with system operation protocols, specified for detailing system functionalities of the novel dNAS, on different wine record management process. The system implementation is categorized into two phases, namely initialization phase and data processing phase, covering system implementations, such as consortium formation and decentralized wine record management on dNAS.

\subsection{Initialization Phase}
The initialization phase includes all the necessary system implementation steps needed to be in place before dNAS can be fully functional for wine record management when wine products moving along the supply chain. The initialization phase is consisted of consortium formation processes with only the sample on-boarding process of new node to enterprise consortium covered in this research.

\subsubsection{On-boarding of New Node to Enterprise Consortium}
\begin{figure*}[h]
    \centering
    \captionsetup{justification=centering}
    \includegraphics[width=1\textwidth]{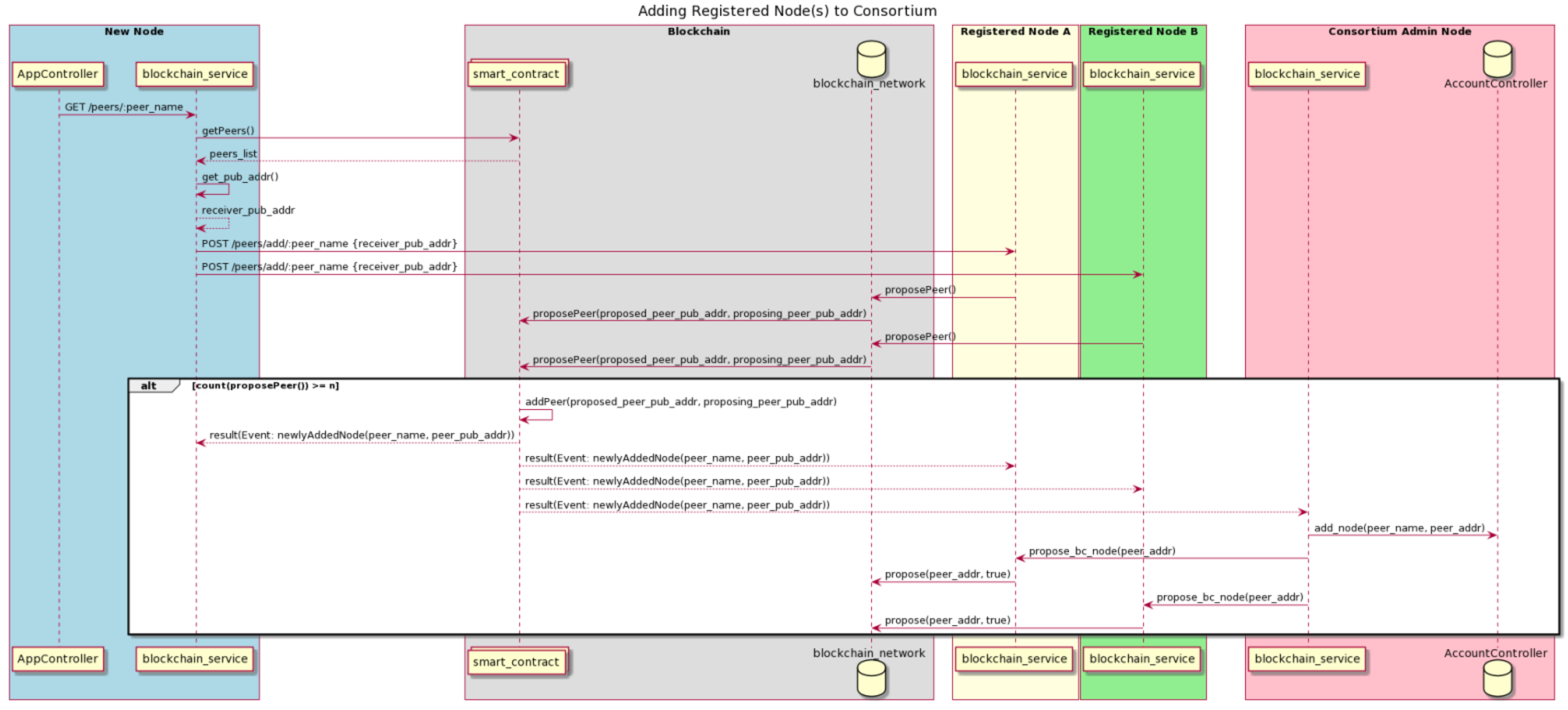}
    \caption{\textit{Adding New Registered Nodes to Enterprise Consortium}}
    \label{fig:onboardingnewnode}
\end{figure*}

After registering for an account to access system components and applications of dNAS hosted by the consortium, such as the database operation web application, and both NFC-enabled mobile applications according to its system role defined, the registered node would then be assigned with the access to every dNAS system component. In order to become a consortium member, an on-boarding operation is needed to execute with a blockchain node instance assigned, which is hosted by the consortium. As detailed in the use case analysis, only registered nodes of winemaker role and supply chain participant role will be assigned with a blockchain node, an IPFS node and key pairs shared and kept in their own instance of key vault service. While registered nodes of wine consumer role utilize shared instances of these distributed system components hosted by the consortium, for data processing operations, such as the wine record validation and appending process. Wine consumers will consequently need to use \emph{ScanWINE} mobile application to validate a bottled wine with wine record returned, before purchasing the wine product with transaction and supply chain data updated in dNAS. The suggested protocol of adding the new node to the consortium is described in \textit{Fig.~\ref{fig:onboardingnewnode}}.

To have a newly registered node added to the consortium, a democratic voting process is performed amongst the current consortium members. It is assumed that such operation is kicked in after a new node is granted access to dNAS as a registered node, approved by the consortium administrator, and assigned the newly registered node of dNAS with an instance of distributed components, including a blockchain node and an IPFS node. With the \emph{URL} of the blockchain node instance specified, the blockchain service sends querying request to reach to the "\emph{getPeers}" method (not limited by role restriction and so it could be accessed even if the blockchain node is a light node of the blockchain network) of the deployed smart contract to get a full list of peer details of existing consortium members. The blockchain service instance of the proposed node then invokes the propose-to-add endpoint of blockchain service instance owned by every consortium member. 

Once the votes of adding such proposed node to the consortium, have amounted to a half of the consortium’s size in term of the number of participating members, \emph{two consequences} would then happen; (1) the peer details of the proposed node will be added to the on-chain registry of consortium members which is in turn a \emph{white list} of the consortium member so as to enable role restriction on accessing methods of the deployed smart contract for wine record management, and (2) once the process of such on-chain operation is completed, an event of adding the new node to the on-chain registry will then be emitted and broadcast to the blockchain service instance of the consortium administrator based on the event listener enabled in the blockchain service, which would in turn initialize a proposal on the blockchain network of adding the blockchain node of such proposed node to be part of the blockchain network, as well as sending requests to the counterpart of every existing consortium member to agree on the proposal via their own blockchain nodes running on the same blockchain network when it has reached the number amounted to \emph{N / 2 + 1} in case a Proof-of-Authority blockchain network is chosen to be implemented in dNAS.

There is also a so-called \emph{bootstrap stage} applied to the on-boarding process, of which a specific starting number, such as 5, of registered node proposed to join the consortium are not required to go through such democratic voting process. Instead, the proposed nodes will be added to the on-chain registry of peer details directly by the consortium administrator, as well as having their blockchain nodes to be added and run on the network, so as to guarantee there is always enough number of blockchain node running on the blockchain network and enough headcounts of the consortium to make democratic decisions. It is also assumed that the voting mechanism of electing a consortium administrator will not be covered in this research and would be explained as a potential limitation of dNAS to develop on-chain governance steps described in future work of dNAS. Similar to the operation of adding registered node, there is also an operation of removing registered node from the consortium which is not covered in the research given their similarity.

\subsection{Data Processing Phase}
The data processing phase includes all the necessary system operations regarding the decentralized wine record management along the supply chain, such as creation, validation and appending, involving different consortium members, with only the wine record creation process and wine record validation process covered in this research necessarily.

\subsubsection{Wine Record Creation Process}
\begin{figure*}[h]
    \centering
    \captionsetup{justification=centering}
    \includegraphics[width=0.8\textwidth]{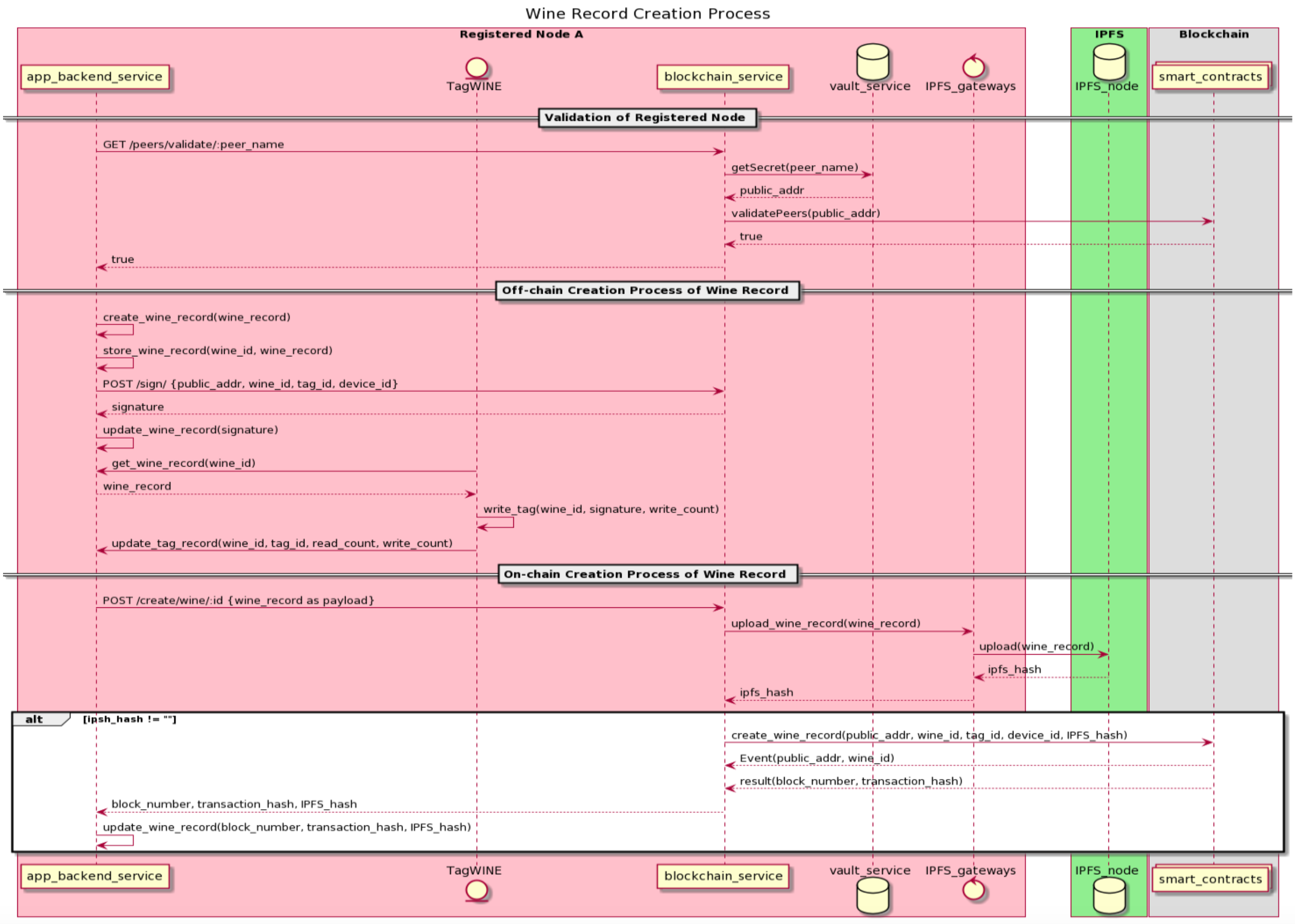}
    \caption{\textit{The Wine Record Creation Process}}
    \label{fig:winerecordcreationprocess}
\end{figure*}

Once a registered node of winemaker role is added to the consortium with its blockchain node being part of the blockchain network, the wine record creation process will be started after the physical bottled wine is produced, packaged and ready to be shipped at the winemaker point, given the fact that the wine record creation process could only be performed by registered nodes of winemaker role, such as the one performed by the Registered Node A, as described in \textit{Fig.~\ref{fig:winerecordcreationprocess}}.

The wine record creation process starts with having the app-backend service, as described in the proposed dNAS system architecture of the Registered Node A, firstly sent a request to its blockchain service instance, by invoking the endpoint of "\emph{/peer/validate}", to confirm if the Registered Node A is indeed a part of the consortium based on the on-chain validation with its public key, stored in its own instance of key vault service, against the version stored in the on-chain registry. Once it is assured that such requesting node is part of the consortium, its app-backend service instance would then be notified and commencing the off-chain creation process of a wine record, including the processes of creating the actual wine record and storing the full wine record to the database of app-backend service.

Invoking an endpoint of signing operations in the blockchain service would produce and return a signature on a \emph{byte-array} of the wine identifier, tag identifier and device identifier involved in the process. In order to write the required data to an NFC tag, Register Node A will then utilize the NFC-enabled \emph{TagWINE} mobile application for which only registered nodes of winemaker role could have access to according to the use cases analysis, to write the wine identifier, signature and the value of write counter to the tag itself. The "\emph{wine_status}" and "\emph{supply_chain_data}" are then updated for which the former will have its updated data fields with latest values, while new entries would be created for the latter whenever there is new activity of any interaction with the tag. The protection mode, with the password predefined randomly and automatically injected by the app-backend service, is also applied to NTAG 216 Ferrite NFC tag once the tag-writing process is completed.

After completing all the required processes of the off-chain creation operation, the on-chain creation process of a wine record is initiated. The app-backend service will invoke the \emph{create} endpoint of the blockchain service, with the subset version of the wine record, instead of the full wine record, as the payload of the API call, and perform operations required on both IPFS network and the blockchain network. Once such wine record subset is uploaded to the IPFS network via a designated IPFS node, a \emph{46-byte content hash} based on \emph{SHA256 hash algorithm} is then returned to the blockchain service. The content hash of IPFS alongside the public address of the registered node and the unique identifiers of the wine product, tag and device are all included in the payload when invoking the "\emph{create_wine_record}" method of a deployed smart contract with transactions sent to the blockchain network. 

The state of the on-chain storage of all these payload fields are updated when the transaction is validated and packaged in the block. The new block is mined by one of the blockchain nodes running on the network, with the respective block number, transaction hash and the event of "\emph{create_wine_record}" is being emitted on-chain and returned to the blockchain service. Both blockchain number and transaction hash are then returned to the app-backend service, updated to supply chain data and transaction data of the full wine record stored in the dedicated database of app-backend service. The corresponding physical bottled wine product is therefore ready to be shipped downstream to the next participant nodes along the supply chain, which is already registered with dNAS to become a part of the enterprise consortium.

\subsubsection{Wine Record Validation Process}
\begin{figure*}[h]
    \centering
    \captionsetup{justification=centering}
    \includegraphics[width=0.8\textwidth]{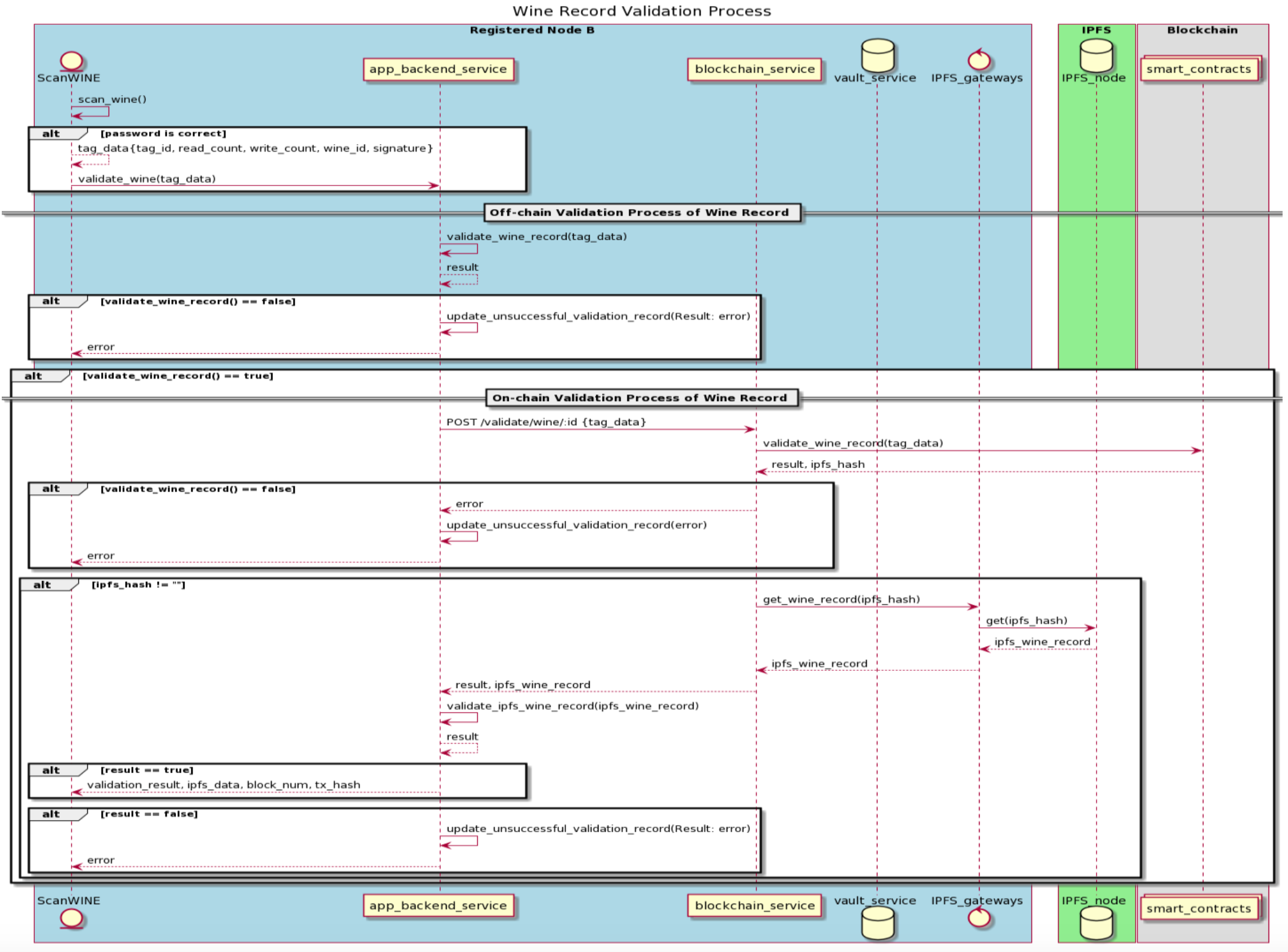}
    \caption{\textit{The Wine Record Validation Process}}
    \label{fig:winerecordvalidationprocess}
\end{figure*}

Once a wine record is created off-chain and on-chain by the registered node of a winemaker role with the tag-writing process completed, those bottled wine products are then shipped to the next node along the supply chain. Assumed the next node along the supply chain is a registered node of dNAS as well as a member of the consortium, such as the Registered Node B described in \textit{Fig.~\ref{fig:winerecordvalidationprocess}}, the corresponding wine record of an incoming wine product would then be brought into a validation process performed on its wine record before the wine product itself could be accepted or purchased, via the NFC-enabled \emph{ScanWINE} mobile application.

The registered nodes of supply chain participant role and wine consumer role will then scan the NFC tags with \emph{ScanWINE} to update the state of that wine record. \emph{ScanWINE} is the only entry point for registered nodes of both roles to transit the state of a wine record, unlike the registered nodes of winemaker role which could utilize the database-operating web application to create and manage wine records. Such wine record validation process is a \emph{three-layered validation process}, namely against (1) the data stored in the database of app-backend service, (2) the on-chain storage, and (3) the updated wine record subset stored on the IPFS network.

When a registered node logged into \emph{ScanWINE} on its device and scan the NFC tag of an incoming bottled wine product, it will automatically get through the password protection applied to the NFC tag as the static password is always injected during the tag-reading process. Data written to the tag, mentioned in the creation process, will then return to \emph{ScanWINE} after which it could invoke the method of validating the wine record based on the data retrieved from the NFC tag. Off-chain validation process against the wine record will then take place to compare the retrieved tag data and those stored in the dedicated database of a shared instance of app-backend service. An error, either classified as (1) modification attack in case the wine identifier and the signature are inconsistent with their counterparts stored in the database, (2) reapplication attack in case either of the "\emph{write_counter}" or "\emph{read_counter}" is different from that of the wine record stored in the database, or (3) cloning attack in case an inconsistent tag identifier is found, would be returned to the app-backend service and consequently updated the "\emph{unsuccessful_validation_data}" of the corresponding wine record, with an error message shown on the user interface of \emph{ScanWINE} application.

The on-chain validation processes of that wine record will therefore be initiated, provided that the off-chain one completed successfully. The app-backend service will send a request to the blockchain service via its "\emph{validate}" endpoint which will further invoke the "\emph{validate_wine_record}" method of the deployed smart contract, named "\emph{wine_data_contract}", with "\emph{tag_data}" as the payload. The public address of the last registered node handling the wine product, which was involved in producing the signature from "\emph{tag_data}", is then recovered on-chain with the identifiers of both the wine product and corresponding tag of "\emph{tag_data}" also supplied. The recovered public address is then validated against its counterpart stored on-chain under the same wine identifier in a dedicated mapping storage on the smart contract.

Similar to the errors returned during the off-chain validation process, an error, either classified as (1) modification attack in case both wine identifier of the payload and recovered public address could not be found or matched on-chain, (2) reapplication attack in case either of the "\emph{write_counter}" or "\emph{read_counter}" is inconsistent with its counterparts stored on-chain, or (3) cloning attack in case an inconsistent tag identifier is found on-chain, would be returned to the app-backend service and updated the respective "\emph{unsuccessful_validation_data}" of a corresponding wine record, with an error message shown on the user interface of \emph{ScanWINE} application. Once the on-chain validation process is successful, the result and latest "\emph{IPFS_hash}" of the specific wine identifier is then returned to the blockchain service, and the latest version of the wine record subset stored in the IPFS network could then be retrieved with the "\emph{IPFS_hash}" specified via its IPFS node. The wine record subset retrieved from IPFS network is then compared with its counterpart stored in the database interacted with the app-backend service. The validation result is then returned to the app-backend service and the user interface of \emph{ScanWINE}.

With a successful validation on the wine record, the overview page of the corresponding wine record is then brought up on the user interface of \emph{ScanWINE} with wine pedigree data, wine status data and some transaction data fields, such as the unique transaction hash with a corresponding block number with which the registered node can query and confirm the latest block and transaction related to this wine product on a blockchain explorer connected to the blockchain network. The "\emph{read_count}" of the tag is incremented after every tag-reading process and so its counterparts stored on-chain and off-chain will also be incremented during every validation operation. The faulty and counterfeited wine product, with an error of any classified attack logged in the "\emph{unsuccessful_validation_data}" of its wine record, should therefore be rejected and returned to its winemakers or the previously registered nodes of the wine products for further operations.

\section{Developing the dNAS Prototype}
Given the system architecture of dNAS elaborated and the system implementation with operational steps explained, the working prototype of dNAS is therefore developed. The development process of dNAS has followed the standard Agile software development methodology involving software project management tools, software development environments, and configurations utilizing modern concepts, such as containerized applications, service-oriented architecture, source version control, cloud environment deployment, etc. Tools and concepts applied to the development process of the individual components in dNAS, such as the blockchain network and smart contracts, are also explained. 

According to the system architecture of dNAS, decentralized system components of dNAS are actually consisted of a blockchain network, smart contracts, an IPFS network and a blockchain service. This chapter includes the implementation details and the design concepts of individual decentralized system components of dNAS.

\subsection{The dNAS Blockchain Network}
The Ethereum developer community has developed many different client implementations, such as Go-Ethereum PoA Clique, Parity PoA Aura and Pantheon PoA, of its options on consensus protocols provided, not to mention there are more enterprise blockchain implementations, such as Quorum, Hyperledger and Tendermint, available for the development of permissioned network for this purpose. There are indeed benefits of having multiple Ethereum clients available which are (1) improved resilience to the network enabled by a faster issue detection and correction – more people interpreting the protocol specification the more chance of these errors to be uncovered and detected, and (2) enhanced security – if there is an attack factor or bug in any of the Ethereum implementations, it means the network is usually working fine as there is a bigger diversity of clients available. \emph{Go-Ethereum nodes} are currently the majority of the Ethereum network including different implementations with different consensus algorithms, and dNAS blockchain network is therefore developed based on the Go-Ethereum PoA implementation.

The network implementation started by creating Ethereum accounts for each of the blockchain nodes with a key pair and its key-store file generated during the process. The secret key could be decrypted if required, such as signing transactions, with the password specified. Ethereum accounts are created and assigned to validator nodes, acting as consortium members, intended to be included in dNAS blockchain network.

The genesis file is essentially the recipe of defining the genesis block, also known as the first block of any blockchain network. There is some significant information set in the genesis file for starting the network; for instance, the "\emph{chainId}" specifies the chain those validator nodes are connected to, the "\emph{period}" specifies a block time for which a block is mined regardless an existence of transactions in the time interval specified, and public addresses of Ethereum accounts generated are also put under "\emph{extraData}" to specify these accounts are going to run as validator nodes when the network started, with their public addresses also listed under "\emph{alloc}" in order to be allocated fund in \emph{gwei} once the network is started. The genesis file is then initiated by every validator node intended to run on the same dNAS blockchain network.

A "\emph{geth}" data folder, consisted of "\emph{chaindata}", "\emph{lightchaindata}", "\emph{node states}" and "\emph{transaction RLP (Recursive Length Prefix)}", will then be created under each of the validator nodes after being initialized with the genesis file specified. The folder will serve as a local copy of global states, block states of specific chain and individual transaction states transited in line with the progress of dNAS blockchain network. The gas limit of the genesis block is also set in the genesis file for the network to get started. However, the gas limit of the block-to-mined is determined by the gas used in its parent block, \emph{which is dynamic and not static in nature} according to the one set for the genesis block, specified as "\emph{gasLimit}" in the genesis file. The mechanism set out in Go-Ethereum implementation is to increase the gasLimit if "\emph{parentGasUsed > parentGasLimit * (2/3)}", otherwise lower it and the amount increased/decreased is determined by how far away it could be from "\emph{parentGasLimit * (2/3) parentGasUsed}".

The individual validator nodes are also ready to be started up to form a network with Ethereum node identifiers specified. During a start-up process of a blockchain node, the gas price is set to zero as it is not considered in any blockchain processes performed in dNAS. The APIs of \emph{remote procedure calls} (RPC) are also available via RPC calls by specifying the internet protocol address of the running node or an \emph{internal process call} (IPC) file under the "\emph{geth}" data folder created for each of the running nodes, to consortium members and consortium administrator for any customized node operation. The node operations could be proposing to add new nodes to dNAS blockchain network with the process via the following sample method, invoked by any blockchain node running on the network, as follows:

\[ clique.propose(public\_address, true) \]

With the number of such proposal amounted to be more than \emph{N / 2 + 1}, where N is the number of running node, the proposed public address representing a proposed node is then added and run as part of dNAS blockchain network. The blockchain node could further be attached to any open-source blockchain explorer, such as \emph{BlockScout}, with the "\emph{ETHEREUM_JSONPRC_HTTP_URL}" specified based on the \emph{provider URL} of the blockchain node.

\subsection{The Decentralized Functional Smart Contracts}
\begin{figure}[h]
    \centering
    \captionsetup{justification=centering}
    \includegraphics[width=0.38\textwidth]{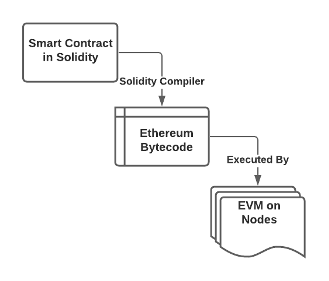}
    \caption{\textit{The Workflow of Ethereum Smart Contract Source}}
    \label{fig:ethereumsmartcontract}
\end{figure}

Smart contracts, hosting functional methods, have the core functional code become fully decentralized and delegated to all the trusted authorities of the consortium and its blockchain network. Every blockchain node on dNAS blockchain network must agree on a current state of the smart contract storage. According to the smart contract mechanism and workflows, such as the source code compiled with Solidity compiler and run by the EVM of every blockchain node of the network, as described in \textit{Fig.~\ref{fig:ethereumsmartcontract}}. 
The smart contract source code, deployed to the blockchain network, is implemented in Solidity with the Truffle framework as its development environment and automated testing library. There are typically three smart contracts developed in dNAS, so as to cover every operation defined in the initialization phase and data process phase, namely (1) the "\emph{WineDataContract}" for wine data operations, (2) the "\emph{PeerRegistryContract}" for the consortium registry management, and (3) the "\emph{Proxy}" which enables a proxy pattern for the purpose of smart contract upgradability.

In order to perform those operations defined under the data process phase, design patterns of mapping storage are defined in smart contracts with individual data types also specified, such as the "\emph{WineDataHash}" is indeed the content hash (\emph{Content ID}) obtained from any IPFS process representing a specific wine record subset mapped to a unique wine identifier which could also be further mapped per different write count as "\emph{iteration}". Similar design patterns of mapping storage could be found applying to the public key of the previous processing node, which is also a registered node of dNAS along the supply chain, and applying to identifiers of different tags and devices. The "\emph{currentImplementation}" and "\emph{initializeCounter}" are in place according to the proxy pattern, which are for checking if a contract address has already been initialized previously.

Events are also defined in the "\emph{WineDataContract}" and they will be emitted when on-chain process of the wine record creation or appending is completed. The events emitted in either of the operation are then captured by the event listener defined in blockchain service. The creation process will include both the public address and hashed device identifier creating the wine record in its event, while the appending process will include both previous and current public address of the wine record.

Moving on with the definition of individual methods, according to the wine record creation process, a specified wine identifier is checked and validated if there is any wine record of such wine identifier has already been logged on-chain previously with the "\emph{require}" statement. The individual variables mapped to the wine identifier are then updated with parameters including a content hash obtained from IPFS node (\emph{WineDataHash}), the public address, tag identifier and device identifier, of the request payload. The on-chain wine record creation process in \textit{Algorithm~\ref{algocreateOnchain}}, is completed by incrementing the write count so as to be matching its counterpart stored in the NFC tag, with an event of the creation process also emitted.

\begin{algorithm}
\caption{Creating On-chain Wine Record Entry}
\label{algocreateOnchain}
\SetAlgoVlined
\SetKwInOut{Input}{Input}\SetKwInOut{Output}{Output}
\Input{wineId, wineDataHash, newPublicAddress, tagId, deviceId}
\Output{Boolean result of create operation}
\BlankLine
    \eIf{MapWriteCount(wineId)==0}{
        $MapPubAddr(wineId)\leftarrow newPublicAddress$\;
        $MapDataHash(wineId)\leftarrow wineDataHash$\;
        $MapTagId(wineId)\leftarrow tagId$\;
        $MapDeviceId(wineId)\leftarrow deviceId$\;
        $MapWriteCount(wineId)+=1$\;
        emit $event$\;
        return $true$\;
    }{
        return $(err, errMessage)$\;
    }
\end{algorithm}

In case there is a wine record stored on-chain for a specific wine identifier with the corresponding physical wine product moved downstream to specific supply chain participants, the wine record is required to be validated, with the wine record validation processes previously detailed. Regarding the on-chain validation of a specific wine record, it could be categorized into two parts, applied to both wine record hash and the signature passed in which are also stored in the NFC tag. For the former, \textit{Algorithm~\ref{algovalidateOnchain}} is a straightforward comparison between the content hash of a specific wine record subset stored in the IPFS network, and its counterpart stored on-chain. It proves the content of the wine record was not altered along the supply chain to the point of validation takes place.

\begin{algorithm}
\caption{Validating the Wine Record (IPFS hash)}
\label{algovalidateOnchain}
\SetAlgoVlined
\SetKwInOut{Input}{Input}\SetKwInOut{Output}{Output}
\Input{wineId, wineDataHash}
\Output{Boolean result of validation}
\BlankLine
    \eIf{MapWriteCount(wineId)!=0}{
        return $MapDataHash(wineId) == wineDataHash$\;
    }{
        return $(err, errMessage)$\;
    }
\end{algorithm}

The latter is to validate the signature, with the sample on-chain method demonstrated in \textit{Algorithm~\ref{algovalidateSig}}, which was produced by the previous processing node with its private key on the data such as the wine identifier, tag identifier and the device identifier, stored in physical NFC tags. The idea of validation on the signature is to recover the public key of the signature on-chain and to validate if the public address recovered is consistent with its counterpart stored under the specific wine identifier on-chain. 

Due to the fact that both tag identifier and device identifier are also retrieved from the on-chain mapping storage with the wine identifier as a key, during the on-chain validation process, both tag identifier and device identifier are also validated to ensure they are consistent with the version of the wine record, stored in the off-chain database, the IPFS network and in the physical NFC tag as the constituents to produce the signature. Both methods for validation can confirm if there has already been an entry of a specific wine identifier stored on-chain to make sure the specific wine record of the wine identifier has already been created on-chain. Adding the \emph{prefix} makes the calculated signature, with the \emph{keccak256} hash function, recognisable as an \emph{Ethereum-specific signature}.

\begin{algorithm}
\caption{On-chain Signature Validation}
\label{algovalidateSig}
\SetAlgoVlined
\SetKwInOut{Input}{Input}\SetKwInOut{Output}{Output}
\Input{wineId, v, r, s}
\Output{Boolean result of validation}
\BlankLine
    \eIf{MapWriteCount(wineId)!=0}{
        $prefix \leftarrow 'Ethereum signature prefix'$\;
        $encodedWineData \leftarrow encode(wineId, TagId_w, DeviceId_w) $\;
        $hashedWineData \leftarrow hash(prefix, encodedWineData)$\;
        return $recover(hashedWineData, v, r, s) == MapPubAddr(wineId)$\;
    }{
        return $(err, errMessage)$\;
    }
\end{algorithm}

Once the validation steps completed successfully, the registered nodes are provided with options to accept a wine product and in case the wine product is accepted, its wine record will be updated with new transaction data and supply chain data appended. The method named "\emph{appendWineRecord}" is also defined in the smart contract for the on-chain appending process by updating the mapping variables accordingly.

\subsection{Design Patterns of Smart Contract Upgradability}
Deployed smart contracts are the shared functional code in the dNAS blockchain network and available for its designated blockchain nodes to invoke so as to perform different operations requested. Just like any other software code, smart contracts are expected to be upgraded for offering new functionalities, fixing software bugs, etc. If the consortium administrator was to upgrade a deployed smart contract, the most straightforward way is to deploy another smart contract and update each of the individual blockchain service instances of dNAS with a new address of the upgraded smart contract. However, this would require a new contract address specified for every blockchain service instance to restart, not to mention there are multiple blockchain service instances each owned by individual consortium members. The contract storage will also be lost and needed to have the stored states from every state-transitioned operation, performed on the smart contract of the previous version, migrated to the new version of the smart contract in order to sustain on-chain states.

\begin{figure}[h]
    \centering
    \captionsetup{justification=centering}
    \includegraphics[width=0.49\textwidth]{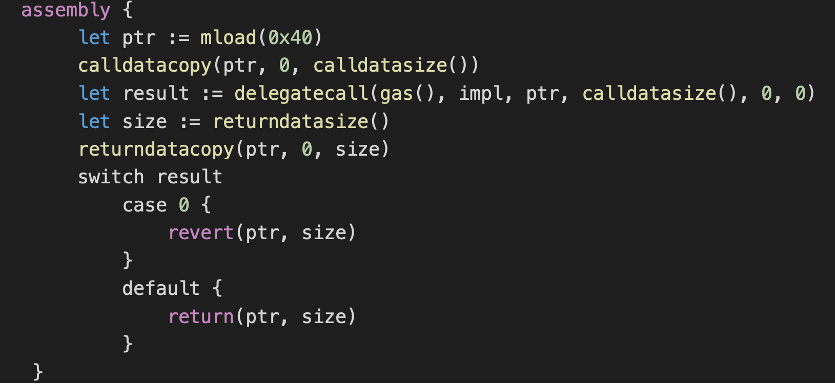}
    \caption{\textit{EVM Assembly Code with delegatecall}}
    \label{fig:delegatecall}
\end{figure}

\begin{figure}[h]
    \centering
    \captionsetup{justification=centering}
    \includegraphics[width=0.46\textwidth]{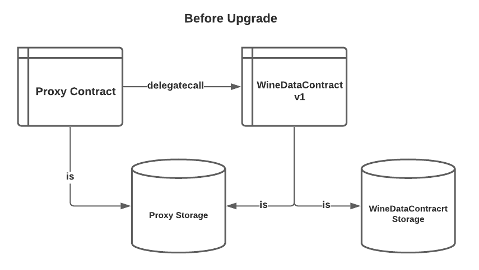}
    \caption{\textit{Proxy Pattern Before Smart Contract Upgrade}}
    \label{fig:proxypatternbefore}
\end{figure}

Given the above drawbacks of any traditional mechanism of upgrading the deployed smart contract, the proxy upgradability pattern offers a common ground regardless of rounds of upgrades with the same proxy contract; nevertheless, it will definitely add a lot of trust to the consortium administrator to perform a smart contract upgrade for the consortium. The proxy contract uses the "\emph{delegatecall}" function of EVM assembly code, as demonstrated in \textit{Fig.~\ref{fig:delegatecall}}, pointing to the contract address of \emph{WineDataContract v1}, as described in \textit{Fig.~\ref{fig:proxypatternbefore}}, alongside the "\emph{msg.sender}" also embedded, which is initiated by the consortium administrator instead of the proxy contract itself.

\begin{figure}[h]
    \centering
    \captionsetup{justification=centering}
    \includegraphics[width=0.46\textwidth]{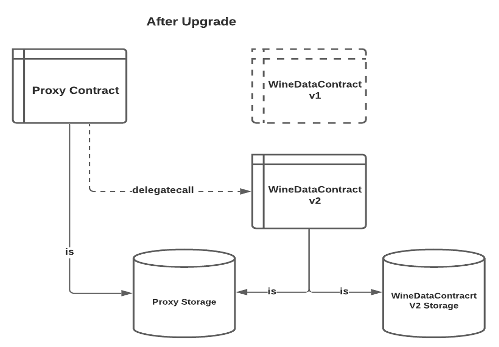}
    \caption{\textit{Proxy Pattern After Smart Contract Upgrade}}
    \label{fig:proxypatternafter}
\end{figure}

The upgrade process is initiated and performed by the consortium administrator, according to the "\emph{upgradeTo}" method defined in the proxy contract. The only thing that the proxy contract needed to operate is to point to a contract address of the upgraded smart contract, which is the \emph{WineDataContract v2} as described in \textit{Fig.~\ref{fig:proxypatternafter}}. 

\subsection{The Blockchain Interface}
The blockchain service is built as an interface between the app-backend service and the blockchain network with other service logic related to the key vault service and IPFS nodes embedded into it, when performing different operations of system implementation. Given the functionalities of the open-source Ethereum interface libraries, such as \emph{Ethers.js} or \emph{Web3.js}, the blockchain service instance is able to invoke methods of the deployed smart contracts, and to interact with its local blockchain node with an object of "\emph{ethers}" or "\emph{web3}" instantiated and contract address specified via injecting environment variables when bringing up the service. Other environment variables represent information, such as the \emph{address} of its IPFS local node as well as both \emph{role_id} and \emph{secret_id} of the key vault service for key management purpose.

\begin{figure}[h]
    \centering
    \captionsetup{justification=centering}
    \includegraphics[width=0.47\textwidth]{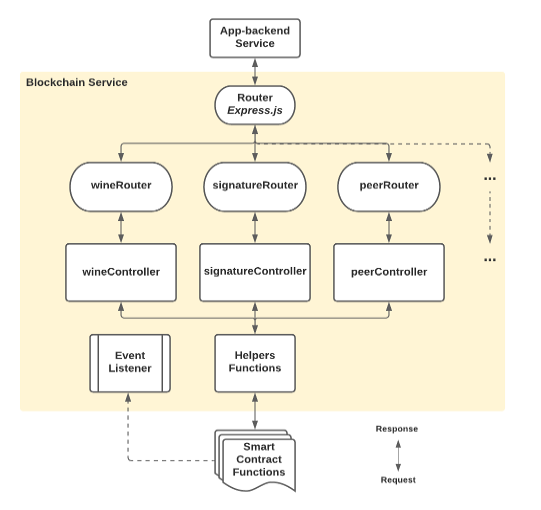}
    \caption{\textit{Functional Request Flow of Blockchain Service}}
    \label{fig:functionalrequestflow}
\end{figure}

The individual routers further link to their respective controllers. The controllers would utilize respective helper functions if some processes are repeated in the same controller functions, such as the process of getting its dedicated IPFS instance and instance of key vault service. Some helper functions would also invoke methods of the deployed smart contracts as shown in the \textit{Fig.~\ref{fig:functionalrequestflow}} demonstrating how the requests from app-backend service could flow through the blockchain service with related functions invoked at different level.

The developed smart contract source code, such as those of \emph{WineDataContract}, \emph{PeerRegistryContract} and \emph{Proxy}, after being compiled as JSON objects, with the Truffle framework, are injected in the blockchain service for further object instantiations and operations.

\subsection{The Decentralized Peer-to-Peer Data Storage with InterPlanetary File System (IPFS)}
As it is too costly and therefore not sensible to store a full version of wine record on-chain. An alternative persistent data source is therefore required, and the IPFS decentralized peer-to-peer data storage is an obvious choice to be deployed in dNAS, under which an immutable and permanent content hash would be obtained from the IPFS network. The IPFS hash of a wine record subset would be stored on-chain under its respective wine identifier, and could be referred whenever there is a request for further operations, such as the wine record validation across different blockchain service instances. In the perspective of security, wine record subsets could not be altered or tampered before being passed to the dNAS blockchain network and the IPFS network, given the fact that every content hash of specific wine record subset, obtained from the IPFS network, will be changed if any of the content of the wine record subset has been updated, which would further lead to an inconsistency with its counterpart stored on-chain when a wine record validation process is performed.

The IPFS functionalities in the blockchain service instance are enabled by "\emph{js-ipfs}". Before performing further IPFS operations, the IPFS node instance is needed to be created in the blockchain service instance. During the wine record creation operation, of which changes are applied to the wine record subsets, the content hash is generated. The content hash generated is with \emph{multi-hash format} and \emph{base-58 encoding} under which the unique multi-hash allows the \emph{exclusivity} of a content hash generated, when the updated wine record subsets are added to the IPFS network. Once the IPFS node instance is generated, the wine record subsets are pushed to the IPFS network with both "\emph{path}" and "\emph{content}" specified in the "\emph{addDataIPFS}" function. The generated content hash of IPFS is then pushed and stored on-chain.

IPFS content hash is retrieved, from the on-chain storage, during the wine record validation operation, and the retrieved content hash is further sent to the IPFS node to get the wine record subset represented, from the IPFS network. When looking up wine record subsets in the IPFS network, the network indeed utilizes the \emph{distributed hash table} (DHT) looking for respective IPFS nodes storing the specified wine record subset behind a content hash. The more times of a wine record subset is retrieved from the IPFS network, the more copies that record existed in the network as the copy will reside on some IPFS nodes of the network, not to mention IPFS nodes with the retrieved data will also act as hosts, accelerating the lookup process from the distributed hash table, for other IPFS nodes looking up the same set of wine record subsets. 

\emph{Object pinning} of retrieved data could also be applied to ensure the data is always available on a local IPFS node, if the retrieved data is needed to be available for more further operations; however, as the retrieved wine record subset is only required for the specific wine record validation operation and so the cache of the local node should be freed up periodically if the data is retrieved with the normal "\emph{cat}" method. IPFS nodes could also be added to or removed from a \emph{private IPFS network} like the one set for dNAS, under which only IPFS nodes owned by any of consortium members and the consortium administrator are part of the private IPFS network according to the node management mechanism described in system implementation. Any message regarding the activities of data addition to or retrieval from the IPFS private network is shared \emph{only} across the network. IPFS nodes on the private network could manage a list of IPFS nodes they have been communicating with, which are also owned by other consortium members of dNAS for \emph{IPFS node management}.

\subsection{The Key Management Module – Key Vault Service}
It is important to know that blockchain transactions could only be sent from an account of a blockchain node, owned by any consortium member, with a valid digital signature produced with private key of the account, such as the digital signature generated during the data processing operation. Given how important the key management of Ethereum accounts to the security model of dNAS, it is therefore vital to keep the private key of any account, involved in dNAS, safe, with a well-defined key management process and selection of key management modules. With the key vault service deployed in dNAS, it could prevent any malicious player from compromising wine record components to negatively impact the data and process integrity of entire dNAS solution, though data stored on-chain is already hashed to obfuscate.

The key vault service of dNAS is developed based on the open-source \emph{Vault by HashiCorp}, which could further be remodelled with \emph{AWS Secrets Manager} or \emph{Akeyless Vault} depending on system requirements, so as to offer functionalities to manage and control access to the secrets of different Ethereum nodes and its accounts respectively, owned by any consortium member, and to manage secrets and contract addresses returned after every smart contract deployment process. Identities are needed to be verified before a specific blockchain service instance could access and interact with a key vault service instance of dNAS. There are two authentication methods available for every distributed key vault services instance, used by any individual consortium member, such as \emph{token-based} and \emph{approle-based}, as defined in "\emph{VaultAuthMethod}". 

The former is based on a unique access token generated by any vault service instance for specific blockchain service, which is also owned by the consortium member. The access token generated is limited by the authentication leases implying that reauthentication is needed after a given lease period so as to continue accessing the key vault service. The latter is about a set of policies and login constraints, such as the \emph{RoleID} and \emph{SecretID} generated, regarding the specific key vault service instance, which must be met to receive a token with those policies set so as to access the key vault service. The distributed blockchain service instance could access to the distributed key vault service instance by specifying the "\emph{VaultAuthMethod}" with necessary metadata, such as the URL of a deployed vault service instance, the API version or both role identifier and secret identifier depending on the authentication method specified, through environment variables when bringing up the instance of distributed key vault service.

Once the key vault service is instantiated with a given blockchain service instance, both owned by a specific consortium member, the secrets of an Ethereum node and account owned by the same consortium member are then ready to be created and stored in the same key vault service instance with a given "\emph{VAULT_PREFIX_KEY}" and \emph{memberID}. According to the smart contract deployment process, consortium administrator is the only role permitted to develop, deploy and upgrade deployed smart contracts, as depicted in use case analysis of dNAS. The contract address of both individual smart contracts and the corresponding proxy contract alongside the public address of the contract owner, which is the consortium administrator for this case, could use a similar design pattern to store the metadata as "\emph{SCDeploymentSecret}" with the corresponding proxy address as a key to the secrets stored.

\section{Conclusion}
With the current challenges on anti-counterfeiting of supply chain industry, the research was performed basing on an implementation-driven research technique with the main question and individual sub-questions listed in the research objectives addressed of this research. A proposed solution, namely the Decentralized NFC-enabled Anti-counterfeiting System (dNAS), is therefore developed to be an autonomous and decentralized application for supply chain anti-counterfeiting and traceability especially for the wine industry, providing a way for registered nodes to verify the legitimacy of wine products with its wine records cryptographically validated. dNAS is appeared to strengthen the capacity in anti-counterfeiting and traceability for wine industry and the wider supply chain industry, while minimising the involvement of centralized parties, such as winemakers of the centralized NFC-enabled anti-counterfeiting system (NAS) for which this research is aimed to decentralize.

The literature analyses of this research cover current alternative resolutions of product anti-counterfeiting, and existing blockchain implementations applied in different industry. The idea of decentralized application was also explained and applied to the development of dNAS as a contribution of this research to explain why ÐApp could exhibit properties, such as zero downtime, immutability and resistance to collusion. The proposed system model design and system architecture of dNAS were implemented based on a set of system requirement defined according to the findings of security analysis performed against the legacy NAS. How dNAS is actually implemented, is explained with system implementation procedures categorized mainly into two phases, namely the \emph{initialization phase} including those operations related to consortium registry management and smart contract deployments, and the \emph{data processing phase} including wine record creation and validation operations of dNAS. The technical details of system development, design patterns applied to different system components and the system execution of dNAS working prototype were also covered and elaborated.

It appeared that the developed prototype of dNAS was a solid example demonstrating how a legacy supply chain anti-counterfeiting and traceability system, for wine industry in this case, could be decentralized and reengineered from its centralized architecture which was once hosted and maintained merely by intermediaries. dNAS is built around the idea of enterprise consortium with registered nodes along the supply chain of wine industry, balancing the degree of decentralization, scalability, security and the privacy of the proposed solution, with the concept of enterprise blockchain chosen appropriately as a preferred development model of the blockchain network as well as availability of shared smart contracts with capability to be upgraded with versions shared amongst the consortium members. dNAS will be evaluated with system testing procedures, such as automated testing and performance testing on different system components of dNAS, and the discussion of result from these system testing activities and qualitative system analysis on the prototype of dNAS, will be included in a separate research with limitation, concerns and future opportunities of dNAS also identified.

\bibliographystyle{ieeetr}

\begin{IEEEbiography}[{\includegraphics[width=1.1in,height=1.25in,clip,keepaspectratio]{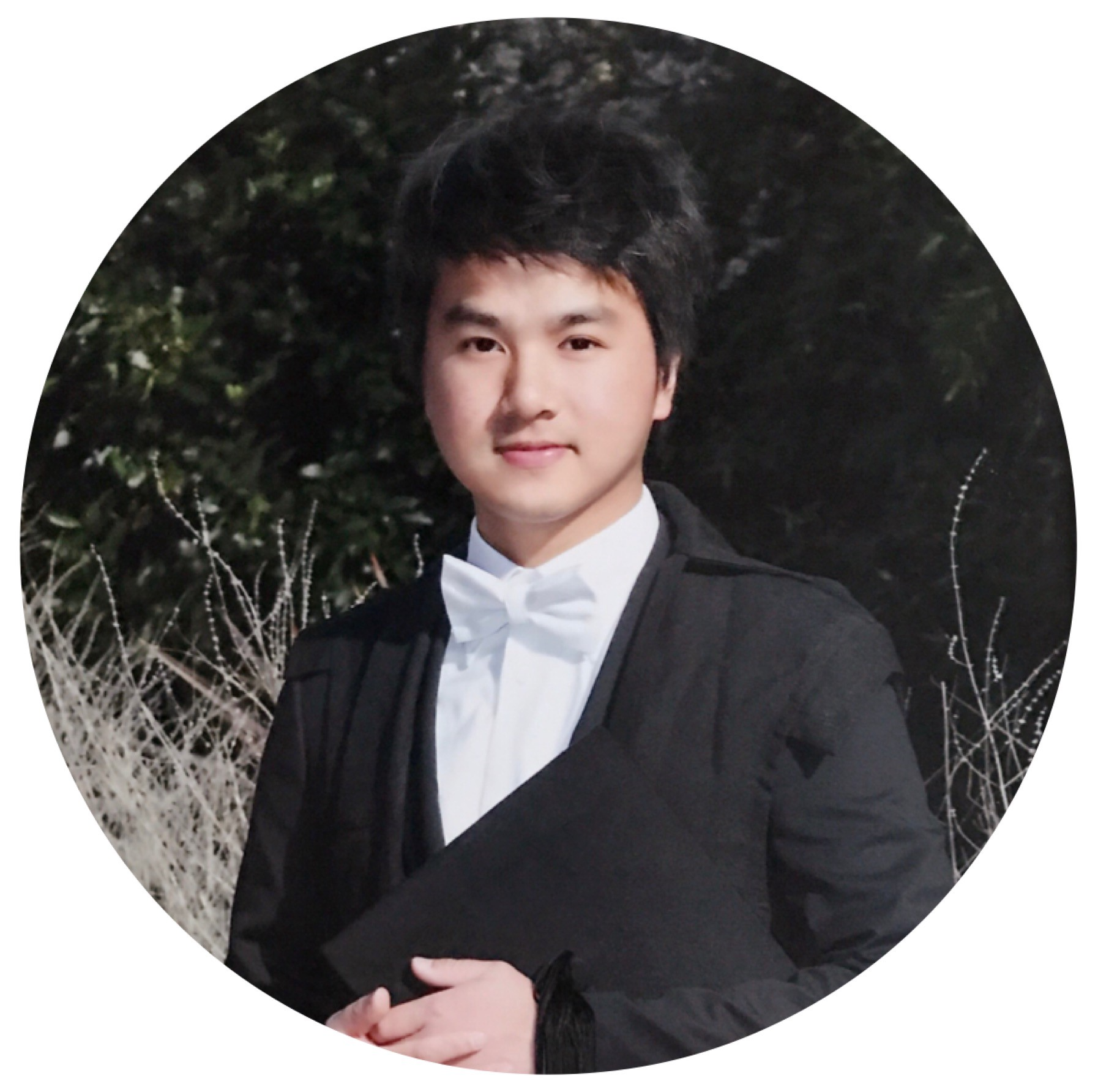}}]{Mr. Neo C.K. Yiu IEEE}
is a computer scientist and software architect specialized in developing decentralized and distributed software solutions for industries. Neo is currently the Lead Software Architect of Blockchain and Cryptography Development at De Beers Group on their end-to-end traceability projects across different value chains with the Tracr™ initiative. Formerly acting as the Director of Technology Development at Oxford Blockchain Society, Neo is currently a board member of the global blockchain advisory board at EC-Council. Neo received his MSc in Computer Science from University of Oxford and BEng in Logistics Engineering and Global Supply Chain Management from The University of Hong Kong.
\end{IEEEbiography}
\vfill

\newpage
\appendices
\section{Background}
\subsection{ The Overview of the Legacy NAS (NFC-enabled Anti-Counterfeiting System)} \label{a1}
\begin{figure*}[h]
    \centering
    \captionsetup{justification=centering}
    \includegraphics[width=0.9\textwidth]{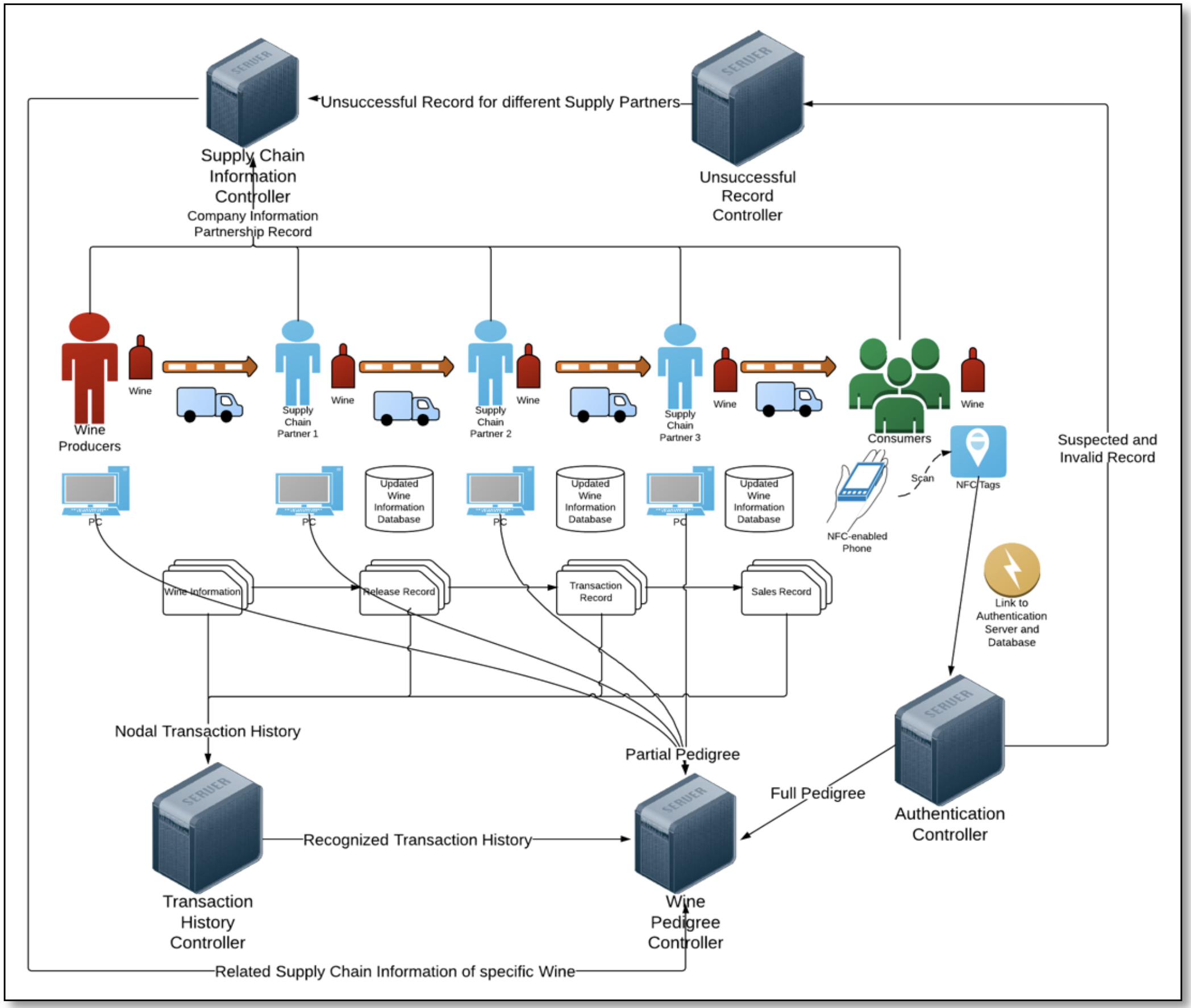}
    \caption{\textit{The System Architecture of Legacy NAS (Source: Neo Y.)}}
    \label{fig:nasarchitecture}
\end{figure*}

NAS with centralised data architectures which are predominantly based on the typical and familiar cloud-based client-server architecture style is demonstrated in \textit{Fig.~\ref{fig:nasarchitecture}}.

The whole NAS solution is consisted of FIVE main components, which are (1) the back-end system with the web-based database management user interface for wine data management performed by winemakers, for which management of data columns of specific wine products which are in their custody can be performed by winemakers, (2) an NFC-enabled mobile application, \emph{ScanWINE}, for tag-reading purpose of wine products at retailer points for wine consumers or supply chain participants of the supply chain before accepting a wine product, (3) another NFC-enabled mobile application, \emph{TagWINE}, performing tag-writing purpose for wine products at wine-bottling stage by winemakers, (4) the NFC tags packaged on the bottleneck for those purposes and actions, and (5) any NFC-enabled smartphones or tablets of supply chain participants and wine consumers along the supply chain.

There are FOUR major categories of data to be processed along the supply chain with NAS, namely the (1) nodal transaction history data, (2) supply chain data, (3) wine pedigree data which is processed with its dedicated controllers based on their predefined schema and data models, and (4) unsuccessful validation data returned from any unsuccessful validation at the stage of accepting wine products. As the wine products being processed and handled by different nodes along the supply chain with the data updated by scanning the NFC tags of the wine product using the tag-reading \emph{ScanWINE} with the state of wine record to be updated accordingly to the database. These categories of data are updated along the supply chain until the point of purchase at which wine consumers use the tag-reading \emph{ScanWINE} to scan the NFC tags and retrieve the data such as wine pedigree data and transaction data for real-time validations to determine if the wine product is counterfeit or not.

A unique identifier is assigned to each wine product and is written into the NFC tag. Such tag-attached wine products are then shipped from winemakers to different nodes along the supply chain. During the transportation process of the wine products along the supply chain, each involved node could interact with the NFC tags and add the aforementioned four categories of data into the NFC tags respectively. In this way, the next node can check whether or not the wine products have already passed through the legitimate supply chain. If any inconsistency is found at any node, such wine products may be considered as counterfeits and should be returned to winemakers. However, once the wine product reached post-purchase stage and circulated in any customer-to-customer markets, its authenticity is no longer guaranteed, as anyone who has an NFC reader could interfere and clone tags' data. Therefore, it is important to develop anti-counterfeiting and traceability systems that could work even when the data of the tag is cloned in the post-purchase supply chain with the security attacks detected and prevented on any potential adversary state transition took place.

\section{System Model Definition}
\subsection{Use Case Analysis Diagram of dNAS} \label{b1}
\begin{figure*}[h]
    \centering
    \captionsetup{justification=centering}
    \includegraphics[width=0.98\textwidth]{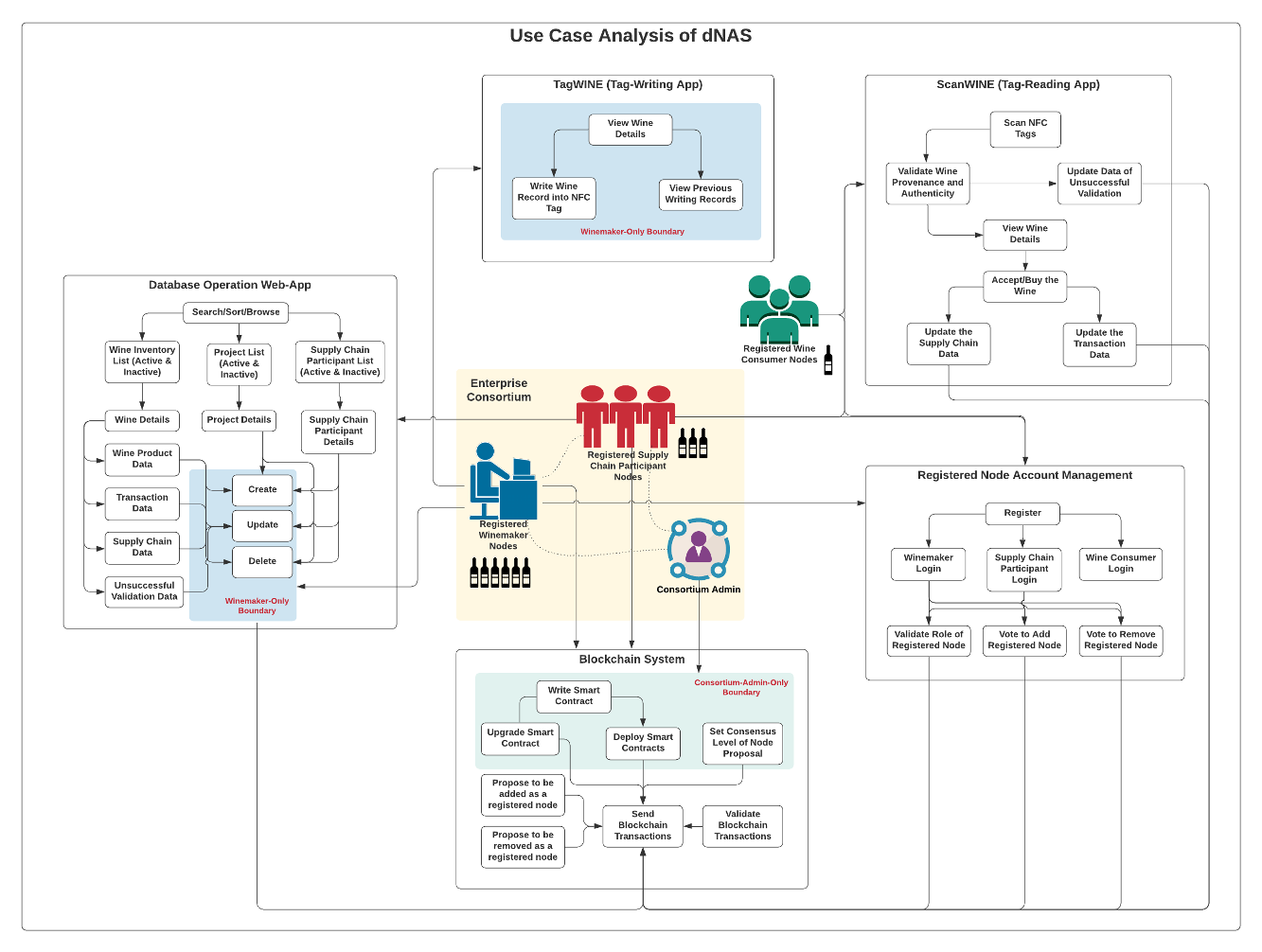}
    \caption{\textit{Use Case Analysis Diagram of dNAS}}
    \label{fig:dnasusecase}
\end{figure*}

The Use Case Analysis Diagram of dNAS is depicted in \textit{Fig.~\ref{fig:dnasusecase}}.

\subsection{System Architecture Diagram of dNAS} \label{b2}
\begin{figure*}[h]
    \centering
    \captionsetup{justification=centering}
    \includegraphics[width=1\textwidth]{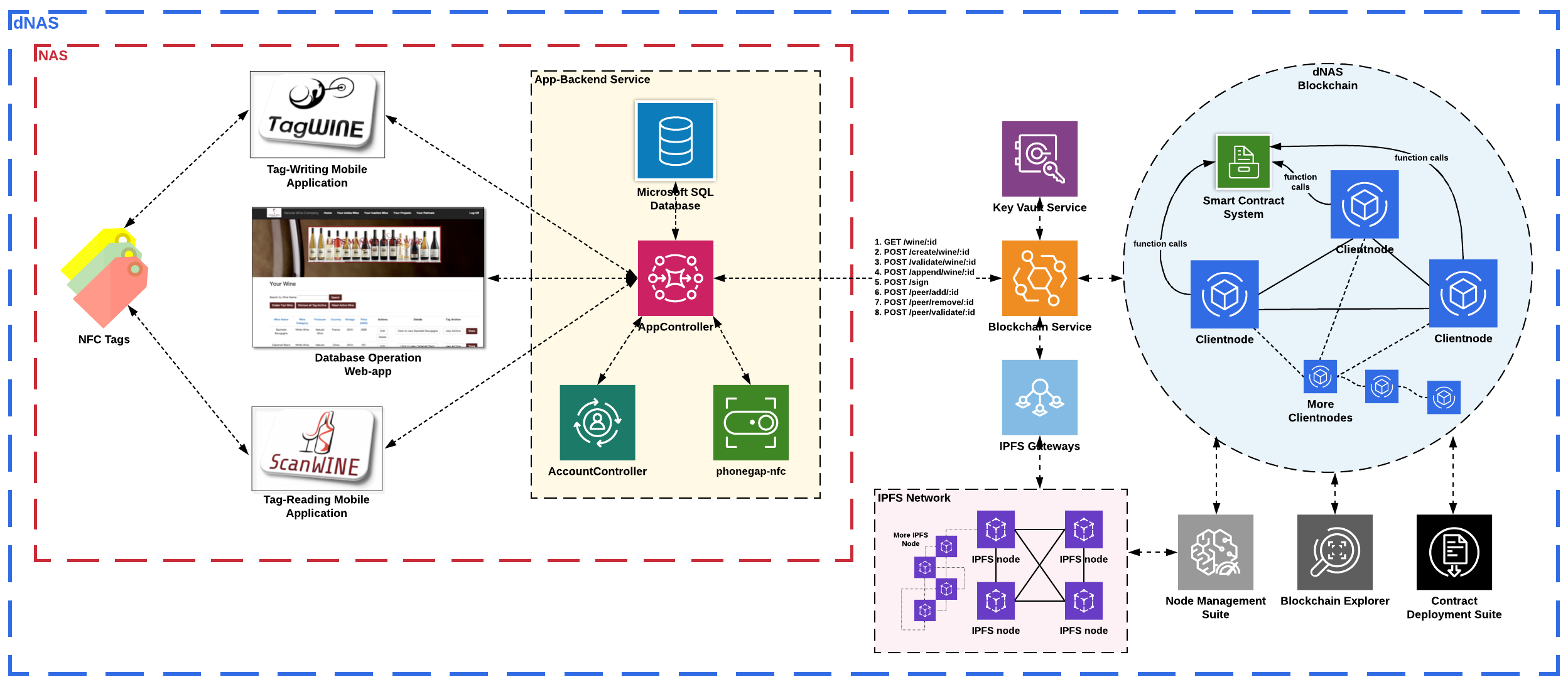}
    \caption{\textit{System Architecture Diagram of dNAS}}
    \label{fig:sysarchi}
\end{figure*}

The Full System Architecture Diagram of dNAS is depicted in \textit{Fig.~\ref{fig:sysarchi}}.

\end{document}